\documentclass[twocolumn,english,prb,showpacs,superscriptaddress]{revtex4}
\usepackage[T1]{fontenc}
\usepackage[latin1]{inputenc}
\usepackage{amsmath}
\usepackage{graphicx}
\usepackage{amssymb}
\usepackage{esint}

\makeatletter

\newcommand{\noun}[1]{\textsc{#1}}
\providecommand{\tabularnewline}{\\}

\@ifundefined{textcolor}{}
{%
 \definecolor{BLACK}{gray}{0}
 \definecolor{WHITE}{gray}{1}
 \definecolor{RED}{rgb}{1,0,0}
 \definecolor{GREEN}{rgb}{0,1,0}
 \definecolor{BLUE}{rgb}{0,0,1}
 \definecolor{CYAN}{cmyk}{1,0,0,0}
 \definecolor{MAGENTA}{cmyk}{0,1,0,0}
 \definecolor{YELLOW}{cmyk}{0,0,1,0}
 }

\@ifundefined{definecolor}
 {\@ifundefined{definecolor}
 {\usepackage{color}}{}
}{}

\renewcommand{\vec}[1]{\mathbf{#1}}
\usepackage{wasysym}

\usepackage{lineno}

\makeatother

\usepackage{babel}\makeatother

\makeatother

\usepackage{babel}

\makeatother

\usepackage{babel}

\begin{document}

\title{Precise response functions in all-electron methods: Application to
the optimized-effective-potential approach}

\author{Markus Betzinger}

\email{m.betzinger@fz-juelich.de}

\affiliation{Peter Grünberg Institut and Institute for Advanced Simulation, Forschungszentrum
Jülich and JARA, D-52425 Jülich, Germany}

\author{Christoph Friedrich}

\affiliation{Peter Grünberg Institut and Institute for Advanced Simulation, Forschungszentrum
Jülich and JARA, D-52425 Jülich, Germany}

\author{Andreas Görling}

\affiliation{Lehrstuhl für Theoretische Chemie, Universität Erlangen-Nürnberg,
Egerlandstr.~3, D-91058 Erlangen, Germany}

\author{Stefan Blügel}

\affiliation{Peter Grünberg Institut and Institute for Advanced Simulation, Forschungszentrum
Jülich and JARA, D-52425 Jülich, Germany}
\begin{abstract}
The optimized-effective-potential (OEP) method is a special technique
to construct local Kohn-Sham potentials from general orbital-dependent
energy functionals. In a recent publication {[}M.~Betzinger, C.~Friedrich,
S.~Blügel, A.~Görling, Phys.~Rev.~B 83, 045105 (2011){]} we showed
that uneconomically large basis sets were required to obtain a smooth
local potential without spurious oscillations within the full-potential
linearized augmented-plane-wave method (FLAPW). This could be attributed
to the slow convergence behavior of the density response function.
In this paper, we derive an \textit{incomplete-basis-set correction}
for the response, which consists of two terms: (1) a correction that
is formally similar to the Pulay correction in atomic-force calculations
and (2) a numerically more important basis response term originating
from the potential dependence of the basis functions. The basis response
term is constructed from the solutions of radial Sternheimer equations
in the muffin-tin spheres. With these corrections the local potential
converges at much smaller basis sets, at much fewer states, and its
construction becomes numerically very stable. We analyze the improvements
for rock-salt ScN and report results for BN, AlN, and GaN, as well
as the perovskites $\mathrm{CaTiO}_{3}$, $\mathrm{SrTiO}_{3}$, and
$\mathrm{BaTiO}_{3}$. The incomplete-basis-set correction can be
applied to other electronic-structure methods with potential-dependent
basis sets and opens the perspective to investigate a broad spectrum
of problems in theoretical solid-state physics that involve response
functions. 
\end{abstract}


\maketitle

\section{Introduction}

Density-functional theory (DFT)\cite{Hohenberg-Kohn,DFT-review} in
the Kohn-Sham (KS) formalism\cite{Kohn-Sham} has developed into a
standard method for computational calculations of electronic properties
due to its theoretical and numerical simplicity that goes along with
the required accuracy for a large range of materials. In this theory,
the many-body exchange and correlation effects that make the microscopic
quantum-mechanical description of many-electron systems so complicated
are hidden in a formally simple local potential, the exchange-correlation
(xc) potential. Together with the classical electrostatic potential
created by the electronic and nuclear charges, it forms the local
effective potential for a fictitious system of non-interacting electrons,
the KS system, whose electron density, by construction, coincides
with that of the real system. Physical quantities, such as the total
energy, inter-atomic forces, dipole and magnetic moments, etc., are
then calculated as functionals of the electron density.

However, the mathematical form of the xc energy functional, from which
the xc potential derives, is unknown, and one must resort to approximations
in practice. Surprisingly, already simple approximations, such as
the local-density (LDA)\cite{LDA-Ceperley/Adler,LDA-VWN} and generalized
gradient approximations (GGA),\cite{GGA-PW,GGA-PBE} give reliable
results for a wide range of materials and properties. However, as
the functionals have been applied, over the years, to more and more
complex materials and properties, shortcomings have become apparent:
The spurious self-interaction error, inherent to the LDA and GGA functionals,
leads to an unphysical description of localized states, which as a
result appear too high in energy and give rise to erroneous hybridization
effects. Second, the LDA and GGA xc functionals do not exhibit a derivative
discontinuity at integral particle numbers leading to semiconductor
band gaps that are underestimated by 40\% or more with respect to
experiment.\cite{Band-gap-problem-of-DFT,KS-GAP1} Furthermore, they
show the wrong asymptotic behavior when exchange and correlation effects
of spatially separate but interacting parts of a many-electron system
are investigated.

Orbital-dependent functionals form a new class of xc functionals.\cite{Review:O-functionals,Review-Goerling}
Already the formally simple exact-exchange (EXX) functional,\cite{EXX1,EXX2,EXX3}
which treats electron exchange exactly but neglects correlation altogether,
remedies the aforementioned deficiencies of the more conventional
local or semi-local functionals. It has been shown that the EXX functional
leads to KS band gaps that are in much better agreement with experiment.\cite{EXX-LMTO-Kotani-I,EXX-LMTO-Kotani-II,EXX-PP-Staedele-1,EXX-PP-Staedele-2,EXX-PP-Fleszar,FLAPW-EXX}
By definition, the KS band gaps do not contain the derivative discontinuity,\cite{KS-GAP1,KS-GAP2}
which indicates that the effect of neglecting correlation is roughly
of the same magnitude as the discontinuity but of different sign.\cite{KS-GAP+DIS1,KS-GAP+DIS2}
Also, localized $d$- and $f$-electron states appear at larger binding
energies compared to local or semilocal functionals, which is a result
of the exactly compensated self-interaction error.

The xc potential is given by the functional derivative of the xc energy
functional with respect to the electron density. In the case of the
conventional LDA or GGA functionals, the functional derivative translates
into a derivative of a function and is evaluated straightforwardly.
In the case of orbital-dependent functionals, such as the exact-exchange
(EXX) functional, the construction of the local xc potential is much
more involved, because these functionals only depend implicitly on
the density through the orbitals, and a chain-rule must be applied
to evaluate the functional derivative leading to the optimized-effective-potential
(OEP) equation.\cite{EXX3,Review-Goerling}

This integral equation involves two kinds of response functions, one
for the electron density and one for the KS wave functions, as well
as a nonlocal term, which corresponds to the Hartree-Fock potential
in the case of the EXX functional. The former response function, on
which we will focus in the following, describes the linear response
of the electron density with respect to changes in the KS effective
potential and is usually calculated by standard perturbation theory
as a sum over occupied and unoccupied states.

Thus, in contrast to the Hartree-Fock method, the OEP approach involves
the whole spectrum of unoccupied states. Physically, the unoccupied
states provide variational freedom for the electron density to respond
to the changes of the effective potential. In practical \textit{ab
initio} calculations, one finds that single-particle states up to
high energies have to be taken into account, which calls for a basis
set that is able to yield accurate wave functions over a very wide
energy range, spanning from the occupied region up to high-lying unoccupied
states. This places a much higher demand on the basis than in conventional
LDA, GGA, or hybrid-functional calculations, where only the electron
density must be described accurately, while the OEP approach requires
a sufficiently accurate description of the response functions, as
well. This is a serious issue in all-electron methods using so-called
linearized basis sets that are optimized for a certain energy region
by construction and involve atom-centered numerical basis functions
that are adapted to the local effective potential at pre-defined energies.
Examples are the full-potential linearized augmented-plane-wave (FLAPW)\cite{FLAPW1,FLAPW2,FLAPW3}
and linearized muffin-tin orbital (LMTO)\cite{Local_Orbitals_Andersen,Skriver,Full-potential-LMTO}
method, but also methods that rely on pre-calculated and tabulated
basis sets constructed from atomic calculations.\cite{FHI-AIMS,Siesta,DMOL,OPENMX}
A similar problem arises in pseudopotential approaches due to the
pseudized form of the potential around the atomic nuclei, which yields
accurate states only in a finite energy region.

In a recent publication,\cite{FLAPW-EXX} we reported an implementation
of the EXX-OEP approach within the all-electron FLAPW method.\cite{FLAPW1,FLAPW2,FLAPW3}
We employed the mixed product basis (MPB)\cite{CoulombMatrix-MixedBasis,PBE0-NonLocalExactExchangePotential,GW-MixedBasis}
to reformulate the OEP integral equation in terms of a matrix equation
that can be solved by standard numerical linear algebra tools. This
approach also enables the construction of the local EXX potential
without shape approximations. We found that the basis sets are not
independent: the LAPW basis must be converged with respect to a given
MPB. This balance condition is a direct consequence of the sum-over-states
problem described above. Only a highly converged LAPW basis, which
corresponds to a large number of unoccupied states, lends the density
enough flexibility to react adequately to the changes of the potential,
thus leading to a well converged response function. As demonstrated
in Ref.~\onlinecite{FLAPW-EXX}, compared to conventional LDA or
GGA calculations, the number of basis functions had to be increased
typically by a factor of four or five, which from a practical point
of view, cannot be a viable approach and effectively restricts the
method to very small systems. A similar behavior was observed for
Gaussian basis sets.\cite{Balance-Gaussian}

In this paper, we present a numerical correction for the response
functions, with which the balance condition is achieved with a considerably
smaller LAPW basis leading to much faster and numerically more stable
calculations. Furthermore, much fewer empty bands are needed for the
construction of the response functions. The correction relies on the
observation that the LAPW basis is explicitly potential dependent
and optimized for a given effective potential. Any change in this
potential (other than a mere constant) will render the basis sub-optimal
or even inadequate. One way to deal with this issue is cranking up
the LAPW basis such that it can describe the Hilbert spaces for the
unperturbed and perturbed potentials at the same time -- leading to
large computational costs as we have seen in Ref.~\onlinecite{FLAPW-EXX}.
In this paper, we pursue a different approach, which improves the
precision of the calculated response functions considerably without
the need of using larger basis sets and at the expense of only a small
computational overhead. We derive a numerical correction by taking
explicitly into account the changes of the LAPW basis induced by the
perturbations. These changes directly follow from the potential dependence
of the basis functions and are calculated by solving radial scalar-relativistic
Sternheimer equations in the muffin-tin spheres. Similarly, the response
of the core states obeys fully relativistic Sternheimer equations.
Additionally, we take account of the fact that the eigenfunctions
of a Hamiltonian represented in a finite basis set are not exact eigenfunctions
of the Hamiltonian operator, in general. Both corrections vanish in
the limit of an infinite, complete basis. As they correct for different
aspects of the incompleteness of a finite basis set, we refer to them
as the \emph{incomplete-basis-set correction}.

Similar corrections are employed, for example, when shifts of the
Bloch vector are considered in $\vec{k}\cdot\vec{p}$ perturbation
theory,\cite{LAPW-kp-formalism} when the magnetic moment is rotated,\cite{Faehnle}
and when atomic positions are varied, i.e., in calculations of atomic
forces\cite{LAPW-Forces-I,LAPW-Forces-II} and phonon band structures.\cite{Savrasov,Draxl-Phonons}
While in the former cases the effective potential remains unchanged,
it does change in the latter case in principle, which should give
rise to corresponding variations in the muffin-tin basis functions.
Nonetheless, this effect has always been neglected (rigid basis approximation)
arguing that it is small, and one takes only variations into account
that are related to the spatial displacements of the atom-centered
basis functions.  The potential-dependent variations in the MT basis
functions are thus complementary to the incomplete-basis-set correction
known from force calculations and potentially improve the accuracy
of atomic forces, too. We intend to test this conjecture in a future
work. 

The paper is organized as follows. Sections~\ref{sec: Theory} and
\ref{sec: FLAPW-Method} give a brief introduction into the OEP formalism
and the FLAPW method. After a short recapitulation of the EXX-OEP
implementation, we develop the incomplete-basis-set correction in
detail in Section~\ref{sec: FLAPW-Method}. As a practical example
we will use the EXX functional for the calculations. However, we note
that the numerical corrections for the response quantities are generally
applicable, irrespective of the employed orbital-dependent functional.
The incomplete-basis-set correction facilitates the construction of
the EXX potential considerably, in terms of both computational efficiency
and numerical stability, as we will demonstrate in Sec.~\ref{sec:Performance}
for the case of scandium nitride. In Sec.~\ref{sec: results} we
show results for the nitrides BN, AlN, and GaN, as well as the perovskites
$\mathrm{CaTiO}_{3}$, $\mathrm{SrTiO}_{3}$, and $\mathrm{BaTiO}_{3}$.
Finally, we draw our conclusions in Sec.~\ref{sec: Conclusions}.

\section{Theory\label{sec: Theory}}

The KS formalism of DFT relies on the mapping of the interacting electron
system onto a noninteracting system of KS electrons. These fictitious
particles move in an effective potential $V_{\mathrm{eff}}(\vec{r})$
that is defined in such a way that the electron densities of the two
systems coincide. The equation of motion for the noninteracting KS
electrons thus reads\begin{equation}
\left[-\frac{1}{2}\nabla^{2}+V_{\mathrm{eff}}(\vec{r})\right]\varphi_{n\vec{k}}(\vec{r})=\epsilon_{n\vec{k}}\varphi_{n\vec{k}}(\vec{r})\label{eq: KS DGL}\end{equation}
with the KS wave functions $\varphi_{n\vec{k}}(\vec{r})$ and energies
$\epsilon_{n\vec{k}}$ for the Bloch vector $\vec{k}$ and band index
$n$. Here and in the following we restrict ourselves to the spin-unpolarized
case and use Hartree atomic units unless stated otherwise. The generalization
to the spin-polarized case is straightforward.

The effective potential consists of the classical electrostatic potential
created by the electronic and nuclear charges in the system and the
xc potential\begin{equation}
V_{\mathrm{xc}}(\vec{r})=\frac{\delta E_{\mathrm{xc}}[n]}{\delta n(\vec{r})}\label{eq: local xc potential}\end{equation}
with the xc energy functional $E_{\mathrm{xc}}[n]$. In the case of
LDA (GGA), the energy functional is given as a simple function of
the electron density $n(\vec{r})=2\sum_{n\vec{k}}^{\mathrm{occ.}}|\varphi_{n\vec{k}}(\vec{r})|^{2}$
(and its gradient), and the functional derivative {[}Eq.~\eqref{eq: local xc potential}{]}
is evaluated straightforwardly. However, in the more general case
of an orbital-dependent functional, which is only an indirect functional
of the density, the application of the chain rule for functional derivatives
is requested.\cite{EXX3,Review-Goerling} This leads, after multiplication
with the KS single-particle response function\cite{footnote-I}

\begin{equation}
\chi_{\mathrm{s}}(\vec{r},\vec{r}')=\frac{\delta n(\vec{r})}{\delta V_{\mathrm{eff}}(\vec{r}')}=4\sum_{\vec{k}}\sum_{n}^{\mathrm{occ.}}\varphi_{n\vec{k}}^{*}(\vec{r})\frac{\delta\varphi_{n\vec{k}}(\vec{r})}{\delta V_{\mathrm{eff}}(\vec{r}')}\,,\label{eq: density response}\end{equation}
 to the OEP equation in integral form\begin{eqnarray}
\lefteqn{\int\chi_{\mathrm{s}}(\vec{r},\vec{r}')V_{\mathrm{xc}}(\vec{r}')d^{3}r'}\nonumber \\
 & = & 4\sum_{\vec{k}}\sum_{n}\int\frac{\delta E_{\mathrm{xc}}}{\delta\varphi_{n\vec{k}}(\vec{r}')}\frac{\delta\varphi_{n\vec{k}}(\vec{r}')}{\delta V_{\mathrm{eff}}(\vec{r})}d^{3}r'\,,\label{eq: OEP equation}\end{eqnarray}
 where the sums over $\vec{k}$ and $n$ run over all states. Solving
this equation yields the optimized local xc potential $V_{\mathrm{xc}}(\vec{r})$.
We note that additional terms occur on the right-hand side if $E_{\mathrm{xc}}$
exhibits an explicit dependence on the KS eigenvalues, as well. 

In this work, we will employ the EXX functional\begin{eqnarray}
E_{\mathrm{x}} & = & -\sum_{\vec{k},\vec{q}}\sum_{n,n'}^{\mathrm{occ.}}\iint\nonumber \\
 &  & \times\frac{\varphi_{n\vec{k}}^{*}(\vec{r})\varphi_{n'\vec{q}}(\vec{r})\varphi_{n'\vec{q}}^{*}(\vec{r}')\varphi_{n\vec{k}}(\vec{r}')}{|\vec{r}-\vec{r}'|}d^{3}r\, d^{3}r'\end{eqnarray}
 as a practical example, whose functional derivative with respect
to the wave functions yields the expression\begin{equation}
\frac{\delta E_{\mathrm{x}}}{\delta\varphi_{n\vec{k}}(\vec{r})}=\int\varphi_{n\vec{k}}^{*}(\vec{r}')V_{\mathrm{x}}^{\mathrm{NL}}(\vec{r}',\vec{r})\, d^{3}r'\end{equation}
 with the nonlocal exchange potential\begin{equation}
V_{\mathrm{x}}^{\mathrm{NL}}(\vec{r}',\vec{r})=-2\sum_{\vec{q}}\sum_{n}^{\mathrm{occ.}}\frac{\varphi_{n\vec{q}}(\vec{r}')\varphi_{n\vec{q}}^{*}(\vec{r})}{|\vec{r}'-\vec{r}|}\,.\end{equation}

\section{FLAPW method\label{sec: FLAPW-Method}}

In the all-electron FLAPW method,\cite{FLAPW1,FLAPW2,FLAPW3} space
is partitioned into nonoverlapping, atom-centered muffin-tin (MT)
spheres and the remaining interstitial region (IR). The tightly bound
core states are completely confined to the spheres and are calculated
by solving the fully-relativistic radial Dirac equation for the spherically
averaged local effective potential $V_{\mathrm{eff},0}^{a}(r)$, where
$r$ is measured from the sphere center at $\vec{R}^{a}$ and $a$
is an atom index.

The valence-electron wave functions are represented by linear combinations
of basis functions\begin{equation}
\varphi_{n\vec{k}}(\vec{r})=\sum_{\vec{G}}z_{\vec{G}}(n,\vec{k})\phi_{\vec{k}\vec{G}}(\vec{r})\label{eq: wave function in APW}\end{equation}
with the reciprocal lattice vectors $\vec{G}$. For the basis functions
$\phi_{\vec{k}\vec{G}}(\vec{r})$, we employ a bi-partitioned representation:\cite{Local_Orbitals_Andersen,Koelling-Arbman}
plane waves in the interstitial region with a momentum cutoff $|\vec{k}+\vec{G}|\le G_{\mathrm{max}}$
and numerical functions $u_{lp}^{a}(r)Y_{lm}(\vec{r})$ in the MT
sphere of atom $a$ with the spherical harmonics $Y_{lm}(\vec{r})$
and a cutoff value $l_{\mathrm{max}}$ for the angular-momentum quantum
number $l$. The two sets of functions denoted by $p=0,1$ are matched
in value and first radial derivative at the MT sphere boundaries to
yield the LAPW basis functions\begin{widetext}

\begin{equation}
\phi_{\vec{k}\vec{G}}(\vec{r})=\left\{ \begin{array}{cl}
\frac{1}{\sqrt{\Omega}}\exp\left[i(\vec{k}+\vec{G})\cdot\vec{r}\right] & \mathrm{if\,}\vec{r}\in\mathrm{IR}\\
{\displaystyle \sum_{l=0}^{l_{\mathrm{max}}}\sum_{m=-l}^{l}\sum_{p=0}^{1}A_{lmp}^{a}(\vec{k},\vec{G})u_{lp}^{a}(|\vec{r}-\vec{R}^{a}|)Y_{lm}(\widehat{\vec{r}-\vec{R}^{a}})} & \mathrm{if\,}\vec{r}\in\mathrm{MT}(a)\end{array}\right.\label{eq: APWs}\end{equation}
 with the unit-cell volume $\Omega$ and the matching coefficients\begin{equation}
A_{lmp}^{a}(\vec{k},\vec{G})=\frac{4\pi}{\sqrt{\Omega}}i^{l}Y_{lm}^{*}(\vec{k}+\vec{G})\exp[i(\vec{k}+\vec{G})\vec{R}^{a}]\frac{(-1)^{p}}{[u_{l1}^{a}(S^{a}),u_{l0}^{a}(S^{a})]}[u_{l\overline{p}}^{a}(S^{a}),j_{l}(|\vec{k}+\vec{G}|S^{a})]\,,\label{eq: APW matching coefficients}\end{equation}
 \end{widetext}where $\overline{p}=1-p$, $S^{a}$ is the radius
of the MT sphere of atom $a$, $j_{l}(x)$ are the spherical Bessel
functions, and the square bracket denotes the Wronskian \begin{equation}
[f(r),g(r)]=f(r)\frac{dg(r)}{dr}-\frac{df(r)}{dr}g(r)\,.\end{equation}
In the following, we restrict ourselves to the non-relativistic equations.
The scalar-relativistic treatment is deferred to Appendix \ref{sec:Scalar-relativistic-equations}.
The radial function $u_{l0}^{a}(r)$ is the solution of the radial
Schrödinger equation \begin{equation}
h_{l}^{a}ru_{l0}^{a}(r)=\epsilon_{l}^{a}ru_{l0}^{a}(r)\label{eq: radial Schroedinger Eq. for u}\end{equation}
with a pre-defined energy parameter $\epsilon_{l}^{a}$ and the radial
Hamiltonian \begin{equation}
h_{l}^{a}=-\frac{1}{2}\frac{\partial^{2}}{\partial r^{2}}+\frac{l(l+1)}{2r^{2}}+V_{\mathrm{eff},0}^{a}(r)\,.\end{equation}
Its energy derivative $u_{l1}^{a}(r)=\partial u_{l0}^{a}(r)/\partial\epsilon_{l}^{a}$
is obtained from\begin{equation}
h_{l}^{a}ru_{l1}^{a}(r)=\epsilon_{l}^{a}ru_{l1}^{a}(r)+ru_{l0}^{a}(r)\,.\label{eq: radial Schroedinger Eq. for u_dot}\end{equation}
 The energy parameters $\epsilon_{l}^{a}$ are typically placed in
the center of gravity of the $l$-projected density of the occupied
states.

The energy derivative $u_{l1}^{a}(r)$ provides for variational freedom
around these energy parameters. While states close to the energy parameters
are thus accurately described, the basis becomes less adequate for
states that are energetically further away, e.g., semicore and high-lying
unoccupied states. For these states, we may extend the LAPW basis
with local orbitals.\cite{Local_Orbitals1,Local_Orbitals2,Local_Orbitals4,Local_Orbitals3}
These are linear combinations of $u_{l0}^{a}(r)$, $u_{l1}^{a}(r)$,
and solutions of Eq.~\eqref{eq: radial Schroedinger Eq. for u},
$u_{lp}^{a}(r)$ and $p\ge2$, with different energy parameters either
fixed at the semicore level or at higher energies for the unoccupied
states. The linear combinations are such that they vanish at the MT
sphere boundary in value and radial derivative. Thus, the local orbitals
are completely confined to the MT sphere and need not be matched to
a plane wave outside.

\section{Implementation}

In Ref.~\onlinecite{FLAPW-EXX}, we showed that the OEP equation
{[}Eq.~\eqref{eq: OEP equation}{]} can be cast into an algebraic
matrix equation if the quantities are formulated in terms of an auxiliary
basis that is designed to represent wave-function products. For this
purpose we introduced the MPB,\cite{CoulombMatrix-MixedBasis,PBE0-NonLocalExactExchangePotential,GW-MixedBasis}
which is built from products of LAPW basis functions, giving rise
to plane waves $\exp(i\vec{G}\cdot\vec{r})/\sqrt{\Omega}$ in the
interstitial region and MT functions $M_{LP}^{a}(r)Y_{LM}(\vec{r})$
in the MT sphere of atom $a$ with cutoff values $|\vec{G}|\le G'_{\mathrm{max}}$
and $L\le L_{\mathrm{max}}$, respectively. (For the present purpose,
the MPB functions are independent of $\vec{k}$ because of the periodicity
of the local potential.) The radial parts $M_{LP}^{a}(r)$ are constructed
from the products $u_{lp}^{a}(r)u_{l'p'}^{a}(r)$ with $|l-l'|\le L\le l+l'$
and also include the atomic EXX potential. We further form linear
combinations of these functions such that they are continuous in value
and radial derivative at the MT sphere boundaries as well as orthogonal
to a constant function. (The latter is necessary to make the density
response function $\chi_{\mathrm{s}}$ invertible.) For mathematical
details, we refer the reader to Refs.~\onlinecite{CoulombMatrix-MixedBasis,GW-MixedBasis,PBE0-NonLocalExactExchangePotential}.
For the present work, we have further incorporated the boundary condition
of zero slope for the MPB functions at the atomic nuclei. This is
motivated by the observation that the local EXX potentials always
show this behavior (cf.~EXX potentials in Refs.~\onlinecite{Zero-Slope-1,Zero-Slope-2,Zero-Slope-3}).
In Sec.~\ref{sec: results} we will demonstrate that this constraint
improves the shape of the EXX potential in the immediate vicinity
of the atomic nuclei.

In this way, the OEP equation {[}Eq.~\eqref{eq: OEP equation}{]}
for the EXX functional can be expressed as \begin{equation}
\sum_{J}\chi_{\mathrm{s},IJ}V_{\mathrm{x},J}=t_{I}\,,\label{eq: algebraic OEP equation}\end{equation}
 where $I$ is used to index the MPB functions, \begin{eqnarray}
\chi_{\mathrm{s},IJ} & = & \iint M_{I}^{*}(\vec{r})\chi_{\mathrm{s}}(\vec{r},\vec{r}')M_{J}(\vec{r}')d^{3}r\, d^{3}r'\label{eq: response matrix}\end{eqnarray}
is the single-particle response matrix, and\begin{equation}
t_{I}=2\sum_{\vec{k}}\sum_{n}^{\mathrm{occ.}}\iint\frac{\delta E_{\mathrm{x}}}{\delta\varphi_{n\vec{k}}(\vec{r}')}\frac{\delta\varphi_{n\vec{k}}(\vec{r}')}{\delta V_{\mathrm{eff}}(\vec{r})}M_{I}^{*}(\vec{r})d^{3}r\, d^{3}r'\label{eq: t vector}\end{equation}
denotes the vector of the right hand side. Inversion of Eq.~\eqref{eq: algebraic OEP equation}
yields the vector $V_{\mathrm{x},J}$. The local exchange potential
is then finally given by\begin{equation}
V_{\mathrm{x}}(\vec{r})=\sum_{J}V_{\mathrm{x},J}M_{J}(\vec{r})\,.\end{equation}

\subsection{Incomplete-basis-set correction\label{sec: FBSC}}

Both the single-particle response function {[}Eq.~\eqref{eq: response matrix}{]}
and the right-hand side of the OEP equation {[}Eq.~\eqref{eq: t vector}{]}
involve the derivative $\delta\varphi_{n\vec{k}}(\vec{r})/\delta V_{\mathrm{eff}}(\vec{r}')$,
which describes the linear response of the wave function $\varphi_{n\vec{k}}(\vec{r})$
with respect to changes of the effective potential. Eqs.~(3), (16),
and (17) show that in our formalism these changes are parametrized
by the MPB functions $\{M_{I}(\vec{r})\}$. We denote the linear response
of the wave function with respect to $M_{I}(\vec{r})$ by\begin{equation}
\varphi_{n\vec{k},I}^{(1)}(\vec{r})=\int\frac{\delta\varphi_{n\vec{k}}(\vec{r})}{\delta V_{\mathrm{eff}}(\vec{r}')}M_{I}(\vec{r}')\, d^{3}r'\,.\label{eq:phi_prime}\end{equation}
According to first-order perturbation theory, $\varphi_{n\vec{k},I}^{(1)}(\vec{r})$
obeys the normalization condition \begin{equation}
\int\varphi_{n\vec{k}}^{*}(\vec{r})\varphi_{n\vec{k},I}^{(1)}(\vec{r})\, d^{3}r=0\label{eq: normalization condition}\end{equation}
and the inhomogeneous differential equation

\begin{eqnarray}
\left[H-\epsilon_{n\vec{k}}\right]\varphi_{n\vec{k},I}^{(1)}(\vec{r}) & = & \left[\epsilon_{n\vec{k},I}^{(1)}-M_{I}(\vec{r})\right]\varphi_{n\vec{k}}(\vec{r})\label{eq: Sternheimer Eq.}\end{eqnarray}
with $\epsilon_{n\vec{k},I}^{(1)}=\langle\varphi_{n\vec{k}}|M_{I}|\varphi_{n\vec{k}}\rangle$
and $H=-\frac{1}{2}\nabla^{2}+V_{\mathrm{eff}}(\vec{r})$. Equation
\eqref{eq: Sternheimer Eq.}, the so-called Sternheimer equation,\cite{Sternheimer}
follows from linearizing Eq.~\eqref{eq: KS DGL} with respect to
changes of the potential. Left-multiplication with the complex conjugates
of all other eigenstates $\varphi_{n'\vec{k}}(\vec{r})$ ($n'\ne n$),
integration over space, and summing over $n'$ yield the well-known
expression

\begin{equation}
\varphi_{n\vec{k},I}^{(1)}(\vec{r})=\sum_{n'(\ne n)}\frac{\langle\varphi_{n'\vec{k}}|M_{I}|\varphi_{n\vec{k}}\rangle}{\epsilon_{n\vec{k}}-\epsilon_{n'\vec{k}}}\varphi_{n'\vec{k}}(\vec{r})\label{eq: solution Sternheimer in WF space}\end{equation}
 and thus \begin{equation}
\frac{\delta\varphi_{n\vec{k}}(\vec{r})}{\delta V_{\mathrm{eff}}(\vec{r}')}=\sum_{n'(\ne n)}\frac{\varphi_{n'\vec{k}}^{*}(\vec{r}')\varphi_{n\vec{k}}(\vec{r}')}{\epsilon_{n\vec{k}}-\epsilon_{n'\vec{k}}}\varphi_{n'\vec{k}}(\vec{r})\,,\label{eq: wf response Sternheimer in WF space}\end{equation}
where the sum runs over the infinite number of eigenstates of $H$.

As the diagonalization of Eq.~\eqref{eq: KS DGL} in a basis representation
yields a whole spectrum of KS wave functions~$\varphi_{n\vec{k}}(\vec{r})$
and energies~$\epsilon_{n\vec{k}}$, comprising the occupied and
a large number of unoccupied states, the response is straightforwardly
calculated using Eq.~\eqref{eq: wf response Sternheimer in WF space},
and one usually employs this equation for a numerical implementation.

However, the number of available wave functions $N$ is limited in
practice -- it cannot exceed the number of basis functions $N_{\mathrm{LAPW}}$
-- so that the sum in Eq.~\eqref{eq: wf response Sternheimer in WF space}
is truncated $\left(n'\le N\le N_{\mathrm{LAPW}}\right)$ leading
to a loss of accuracy. Equation \eqref{eq: wf response Sternheimer in WF space}
then only accounts for that part of the response that happens to lie
in the Hilbert space spanned by the finite number of wave functions.
As a pragmatic solution, one can increase the number of basis functions
and, thus, the number of eigenstates. However, as already mentioned,
we observed that Eq.~\eqref{eq: wf response Sternheimer in WF space}
converges very slowly with respect to the size of the LAPW basis.
Sufficient convergence is attained only with basis sets that are considerably
larger than the standard one used in GGA or LDA calculations (up to
five times larger even for the simple example of diamond), entailing
high computational costs rendering in part the calculations impractical.

The difficulty of convergence can be overcome by two distinct corrections
to the linear response of the wave function: (i) The necessity for
expanding the wave functions, in practice, into a finite, incomplete
basis set implies that the wave functions $\varphi_{n\vec{k}}(\vec{r})$
are not pointwise exact solutions of Eq.~\eqref{eq: KS DGL}. The
fact that $(H-\epsilon_{n\vec{k}})\varphi_{n\vec{k}}(\vec{r})$ does
not vanish identically for all $\vec{r}$ in the unit cell will give
rise to a small correction that formally resembles the Pulay term
in atomic-force calculations. (ii) More importantly, though, the incompleteness
of the basis leads to a neglect of important response effects, in
particular in the MT spheres, where the standard LAPW basis is optimized
for band energies close to the occupied states, while for higher-lying
bands the basis becomes inadequate. (This is generally true for linearized
methods. In the pseudopotential approach, on the other hand, it is
the one-particle potential itself that is constructed for the occupied
states, making it inappropriate for high energies.) In short, the
MT functions of the standard LAPW basis form a poor basis for the
wave-function response. This is not surprising since it should be
easy to find a perturbation out of the many functions $M_{I}(\vec{r})$
that rotates the resulting wave function out of the Hilbert space
spanned by the basis functions. In fact, each of the MPB functions
(except for the constant function) will have this effect to some extent.
Having said this, the question arises whether it is not possible to
let the Hilbert space itself rotate in the same way as the wave function
$\varphi_{n\vec{k}}(\vec{r})$ does so that $\varphi_{n\vec{k}}^{(1)}(\vec{r})$
remains in the \textit{co-rotating} Hilbert space. In other words,
we seek the response of the basis functions subject to a given perturbing
potential, i.e., $\delta\phi_{\vec{k}\vec{G}}(\vec{r})/\delta V_{\mathrm{eff}}(\vec{r}')$.
Indeed, it will turn out that this response is straightforwardly calculated
in the MT spheres.

We start the derivation by letting a perturbation $\delta V_{\mathrm{eff}}(\vec{r}')$
act on the wave function in Eq.~\eqref{eq: wave function in APW},
which formally gives \begin{equation}
\frac{\delta\varphi_{n\vec{k}}(\vec{r})}{\delta V_{\mathrm{eff}}(\vec{r}')}=\sum_{\vec{G}}\left(\frac{\delta z_{\vec{G}}(n,\vec{k})}{\delta V_{\mathrm{eff}}(\vec{r}')}\phi_{\vec{k}\vec{G}}(\vec{r})+z_{\vec{G}}(n,\vec{k})\frac{\delta\phi_{\vec{k}\vec{G}}(\vec{r})}{\delta V_{\mathrm{eff}}(\vec{r}')}\right)\,.\label{eq: functional derivative phi_nk APW}\end{equation}
 (Local orbitals and core states will be discussed later.) The second
term arises from the fact that the basis functions $\phi_{\vec{k}\vec{G}}(\vec{r})$
depend explicitly on the effective potential through Eqs.~\eqref{eq: radial Schroedinger Eq. for u}
and \eqref{eq: radial Schroedinger Eq. for u_dot}. It contains what
we have termed basis-function response above. We will now construct
this second term explicitly and then see how it combines with the
first term.

As the basis functions depend on the potential only in the MT spheres,
their linear response with respect to a change of the potential is
nonzero only within the spheres. Linearizing Eq.~\eqref{eq: APWs}
gives \begin{widetext}\begin{eqnarray}
\hspace{-0.4cm}\phi_{\vec{k}\vec{G},I}^{(1)}(\vec{r}) & =\hspace{-0.15cm} & \left\{ \begin{array}{cl}
0 & \mathrm{if\,}\vec{r}\in\mathrm{IR}\\
\hspace{-0.1cm}{\displaystyle \sum_{lmp}\left[A_{lmp}^{a}(\vec{k},\vec{G})u_{lp,I}^{a(1)}(|\vec{r}-\vec{R}^{a}|)+A_{lmp,I}^{a(1)}(\vec{k},\vec{G})u_{lp}^{a}(|\vec{r}-\vec{R}^{a}|)\right]Y_{lm}(\widehat{\vec{r}-\vec{R}^{a}})} & \mathrm{if\,}\vec{r}\in\mathrm{MT}(a)\,,\end{array}\right.\label{eq: derivative APW}\end{eqnarray}
 \end{widetext}where the quantities $\phi_{\vec{k}\vec{G},I}^{(1)}(\vec{r})$,
$u_{lp,I}^{a(1)}(r),$ and $A_{lmp,I}^{a(1)}(\vec{k},\vec{G})$ denote
the linear changes of the LAPW basis function, the radial function,
and matching coefficients, respectively, in analogy to Eq.~\eqref{eq:phi_prime}.
Here, we restrict ourselves to MT functions $M_{I}(\vec{r})=M_{I}(r)$
with angular momentum $L=0$. (For simplicity, the function $M_{I}(r)$
is already scaled with $Y_{00}(\hat{\vec{r}})=1/\sqrt{4\pi}$.) Of
course, the radial functions $u_{lp}^{a}(r)$ can also respond to
nonspherical perturbations $(L\neq0)$ of the potential. Then, the
linear response consists of a superposition of $|l-L|,\dots,l+L$
functions. We defer this more general and more complicated case to
a later publication and note that the case $L=0$ gives the most important
contribution.

The functions $u_{lp,I}^{a(1)}(r)$ are obtained from linearizing
Eqs.~\eqref{eq: radial Schroedinger Eq. for u} and \eqref{eq: radial Schroedinger Eq. for u_dot},
which yields Sternheimer equations for the radial functions\begin{equation}
\left[h_{l}^{a}-\epsilon_{l}^{a}\right]ru_{l0,I}^{a(1)}(r)=\left[\epsilon_{l,I}^{a(1)}-M_{I}(r)\right]ru_{l0}^{a}(r)\label{eq: Sternheimer u}\end{equation}
 and \begin{multline}
\left[h_{l}^{a}-\epsilon_{l}^{a}\right]ru_{l1,I}^{a(1)}(r)\\
=\left[\epsilon_{l,I}^{a(1)}-M_{I}(r)\right]ru_{l1}^{a}(r)+ru_{l0,I}^{a(1)}(r)\,.\label{eq: Sternheimer u dot}\end{multline}
 The variation of the energy parameter is given by the expectation
value \begin{equation}
\epsilon_{l,I}^{a(1)}=\langle u_{l0}^{a}|M_{I}|u_{l0}^{a}\rangle\,.\label{eq:e_expec}\end{equation}
The scalar-relativistic versions are again deferred to Appendix \ref{sec:Scalar-relativistic-equations}.
The radial inhomogeneous differential Eqs.~\eqref{eq: Sternheimer u}
and \eqref{eq: Sternheimer u dot} are easily solved by integrating
from the origin $r=0$ to the MT boundary $r=S^{a}$. The resulting
special solutions are not uniquely defined since we may always add
the homogeneous solution $u_{lp}^{a}(r)$ or a multiple of it. This
freedom is removed by requiring that $u_{l0}^{a}(r)$ is normalized,
which leads to the additional conditions\begin{equation}
\int dr\, r^{2}u_{l0}^{a}(r)u_{l0,I}^{a(1)}(r)=0\label{eq:normal1}\end{equation}
 and\begin{eqnarray}
\int dr\, r^{2}u_{l0}^{a}(r)u_{l1,I}^{a(1)}(r) & = & \hspace{-0.1cm}-\hspace{-0.1cm}\int dr\, r^{2}u_{l1}^{a}(r)u_{l0,I}^{a(1)}(r)\,.\label{eq:normal2}\end{eqnarray}
 The Eqs.~\eqref{eq:normal1} and \eqref{eq:normal2} together with
Eq.~\eqref{eq:e_expec} ensure that $u_{lp,I}^{a(1)}(r)$ vanishes
for constant variations of the potential.

\begin{figure}
(a)\includegraphics[scale=0.33,angle=-90]{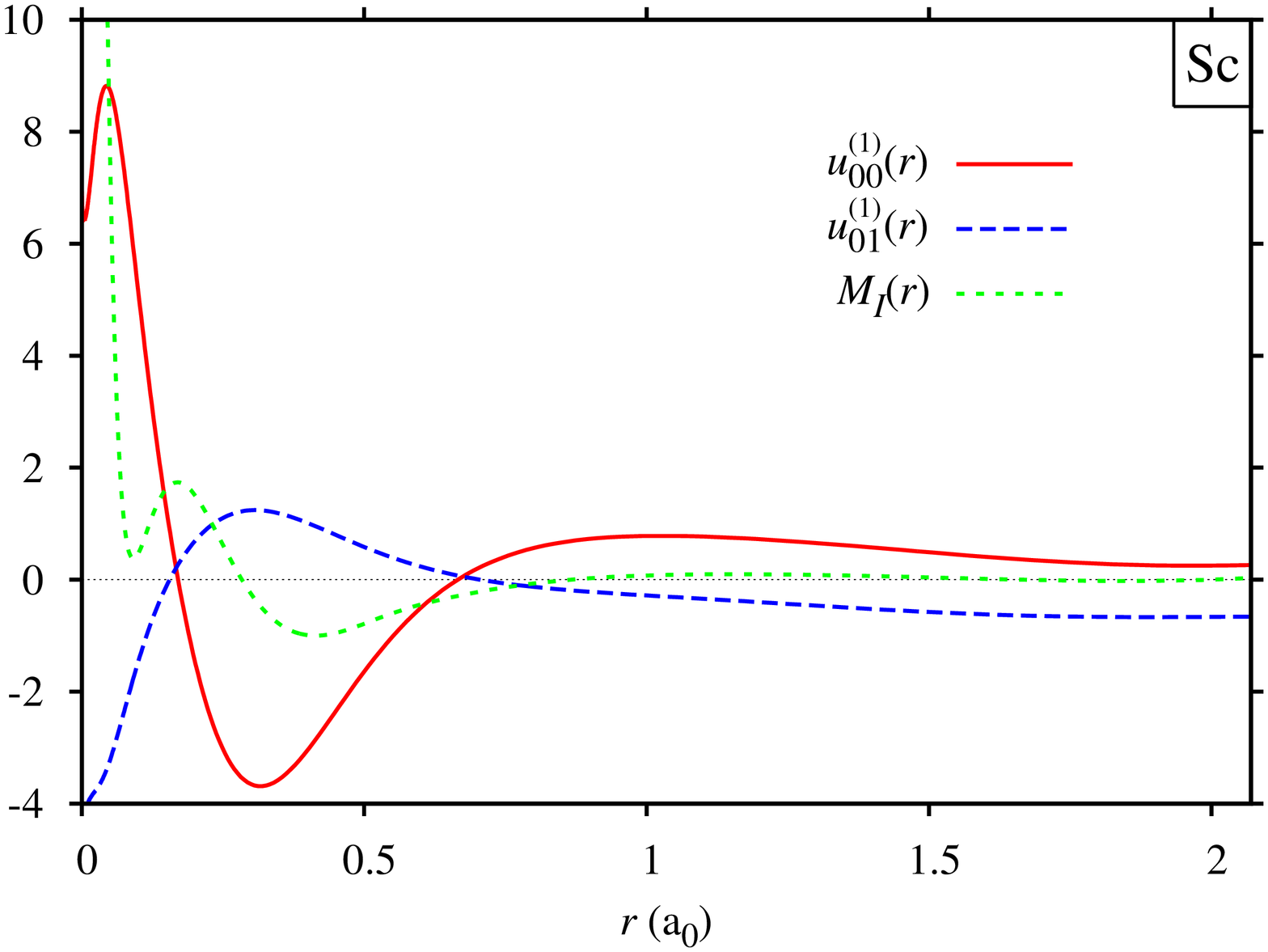}

(b)\includegraphics[scale=0.33,angle=-90]{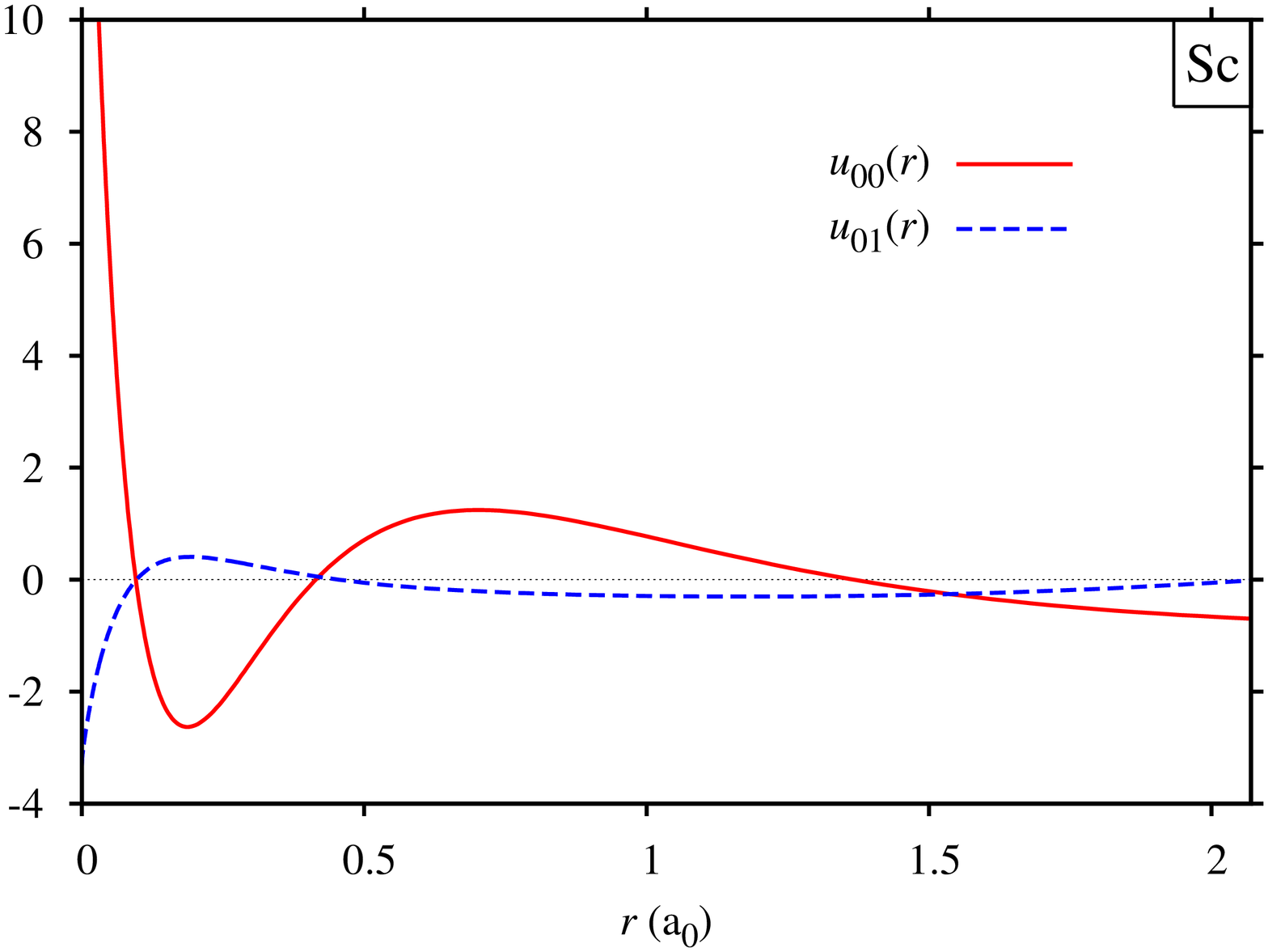}

\caption{(Color Online) (a) Normalized radial basis response functions $u_{00}^{\mathrm{Sc}(1)}(r)$
{[}(red) solid line{]} and $u_{01}^{\mathrm{Sc}(1)}(r)$ {[}(blue)
dashed line{]} as obtained from Eqs.~\eqref{eq: Sternheimer u} and
\eqref{eq: Sternheimer u dot} for angular momentum $l=0$ calculated
in the MT sphere of the Sc atom in rock-salt ScN. The perturbing potential
$M_{I}(r)$ is shown as the (green) dotted line. (b) Corresponding
radial LAPW basis functions $u_{00}^{\mathrm{Sc}}(r)$ {[}Eq.~\eqref{eq: radial Schroedinger Eq. for u}{]}
{[}(red) solid line{]} and $u_{01}^{\mathrm{Sc}}(r)$ {[}Eq.~\eqref{eq: radial Schroedinger Eq. for u_dot}{]}
{[}(blue) dashed line{]}.\label{fig: radial functions}}

\end{figure}

As an example, we show in Fig.~\ref{fig: radial functions}(a) the
radial basis response functions, $u_{00,I}^{\mathrm{Sc}(1)}(r)$ and
$u_{01,I}^{\mathrm{Sc(1)}}(r)$, for the Sc atom of rock-salt ScN
and for $l=0$ as obtained from Eq.~\eqref{eq: Sternheimer u} and
\eqref{eq: Sternheimer u dot} together with the corresponding perturbing
potential $M_{I}(r)$. For comparison the conventional LAPW $s$ functions
$u_{00}^{\mathrm{Sc}}(r)$ and $u_{01}^{\mathrm{Sc}}(r)$ are presented
in Fig.~\ref{fig: radial functions}(b). According to Eq.~\eqref{eq:normal1},
the function $u_{00,I}^{\mathrm{Sc}(1)}(r)$ is orthogonal to $u_{00}^{\mathrm{Sc}}(r)$
and, as obvious from Fig.~\ref{fig: radial functions}(a) and (b),
it is clearly different from $u_{01}^{\mathrm{Sc}}(r)$ by more than
a factor. As a consequence, it lies outside the Hilbert space formed
by the two basis functions and will contribute to the incomplete-basis-set
correction. A similar observation holds for the function $u_{01,I}^{\mathrm{Sc}(1)}(r)$.%

The linear change of the matching coefficients $A_{lmp}^{a}(\vec{k},\vec{G})$
is straightforwardly obtained from differentiating Eq.~\eqref{eq: APW matching coefficients}
with respect to the potential. This finally gives rise to \begin{multline}
A_{lmp,I}^{a(1)}(\vec{k},\vec{G})\\
=(-1)^{p+1}\frac{[u_{l\overline{p}}^{a}(S^{a}),\sum_{p'}A_{lmp'}^{a}(\vec{k},\vec{G})u_{lp',I}^{a(1)}(S^{a})]}{[u_{l1}^{a}(S^{a}),u_{l0}^{a}(S^{a})]}\,.\end{multline}
The coefficients $A_{lmp,I}^{a(1)}(\vec{k},\vec{G})$ guarantee that
the resulting functions $\phi_{\vec{k}\vec{G},I}^{(1)}(\vec{r})$,
as defined in Eq.~\eqref{eq: derivative APW}, and their radial derivatives
continuously go to zero at the MT sphere boundaries.

We note that the rest of the derivation applies generally to spherical
$(L=0)$ and nonspherical perturbations $(L\neq0)$. Once the $\phi_{\vec{k}\vec{G},I}^{(1)}(\vec{r})$
are constructed, linear combinations with the wave-function coefficients
\begin{eqnarray}
\tilde{\varphi}_{n\vec{k},I}^{(1)}(\vec{r}) & = & \sum_{\vec{G}}z_{\vec{G}}(n,\vec{k})\phi_{\vec{k}\vec{G},I}^{(1)}(\vec{r})\end{eqnarray}
yield the second term of Eq.~\eqref{eq: functional derivative phi_nk APW}
for variations that scale with $M_{I}(\vec{r})$ according to Eq.~\eqref{eq:phi_prime}.

The first term of Eq.~\eqref{eq: functional derivative phi_nk APW},
in the following denoted by $\hat{\varphi}_{n\vec{k},I}^{(1)}(\vec{r})$,
lies completely in the Hilbert space spanned by the LAPW basis set.
Accordingly, $\hat{\varphi}_{n\vec{k},I}^{(1)}(\vec{r})$ can be expanded
in terms of the unperturbed KS wave functions\begin{equation}
\hat{\varphi}_{n\vec{k},I}^{(1)}(\vec{r})=\sum_{n'}\langle\varphi_{n'\vec{k}}|\hat{\varphi}_{n\vec{k},I}^{(1)}\rangle\varphi_{n'\vec{k}}(\vec{r})\,.\label{eq: expansion}\end{equation}
The projection coefficient $\langle\varphi_{n'\vec{k}}|\hat{\varphi}_{n\vec{k},I}^{(1)}\rangle$
is obtained by exploiting the fact that $\varphi_{n\vec{k},I}^{(1)}(\vec{r})=\hat{\varphi}_{n\vec{k},I}^{(1)}(\vec{r})+\tilde{\varphi}_{n\vec{k},I}^{(1)}(\vec{r})$
is the solution of Eq.~\eqref{eq: Sternheimer Eq.}. After left multiplication
of Eq.~\eqref{eq: Sternheimer Eq.} with $\varphi_{n'\vec{k}}^{*}(\vec{r})$
($n'\ne n$) and integration over space one yields\begin{multline}
(\epsilon_{n'\vec{k}}-\epsilon_{n\vec{k}})\langle\varphi_{n'\vec{k}}|\hat{\varphi}_{n\vec{k},I}^{(1)}+\tilde{\varphi}_{n\vec{k},I}^{(1)}\rangle+\langle D_{n'\vec{k}}|\hat{\varphi}_{n\vec{k},I}^{(1)}+\tilde{\varphi}_{n\vec{k},I}^{(1)}\rangle\\
=\langle\varphi_{n'\vec{k}}|\epsilon_{n\vec{k},I}^{(1)}-M_{I}|\varphi_{n\vec{k}}\rangle\,,\end{multline}
where we additionally allow for deviations of the calculated wave
functions $\varphi_{n'\vec{k}}(\vec{r})$ from the true eigenfunctions
of the operator $H$ by explicitly treating $D_{n'\vec{k}}(\vec{r})=(H-\epsilon_{n'\vec{k}})\varphi_{n'\vec{k}}(\vec{r})$
as a nonzero quantity. Using $\langle\varphi_{n'\vec{k}}|\varphi_{n\vec{k}}\rangle=0$
and $\langle D_{n'\vec{k}}|\hat{\varphi}_{n\vec{k},I}\rangle=0$ leads
to $\langle\varphi_{n'\vec{k}}|\hat{\varphi}_{n\vec{k},I}^{(1)}\rangle$
for $n\ne n'$. The expansion coefficient for $n=n'$ follows from
the normalization condition Eq.~$ $\eqref{eq: normalization condition}.

By adding up $ $$\hat{\varphi}_{n\vec{k},I}^{(1)}(\vec{r})$ and
$\tilde{\varphi}_{n\vec{k},I}^{(1)}(\vec{r})$ we finally end up with
the instructive result

\begin{multline}
\lefteqn{\varphi_{n\vec{k},I}^{(1)}(\vec{r})=}\\
\sum_{{n'\le N\atop n'(\ne n)}}\left[\frac{\langle\varphi_{n'\vec{k}}|M_{I}|\varphi_{n\vec{k}}\rangle}{\epsilon_{n\vec{k}}-\epsilon_{n'\vec{k}}}+\frac{\langle\varphi_{n'\vec{k}}|H-\epsilon_{n'\vec{k}}|\tilde{\varphi}_{n\vec{k},I}^{(1)}\rangle}{\epsilon_{n\vec{k}}-\epsilon_{n'\vec{k}}}\right]\varphi_{n'\vec{k}}(\vec{r})\\
+\int d^{3}r'\left[\delta(\vec{r}-\vec{r}')-\sum_{n'\le N}\varphi_{n'\vec{k}}(\vec{r})\varphi_{n'\vec{k}}^{*}(\vec{r}')\right]\tilde{\varphi}_{n\vec{k},I}^{(1)}(\vec{r}')\,.\label{eq: wave functions response inkl FBC}\end{multline}
The first term contains the usual expression from first-order perturbation
theory {[}Eq.~\eqref{eq: solution Sternheimer in WF space}{]} and
a correction that takes into account that the wave functions are not
exact eigenfunctions of the Hamiltonian operator $H$ due to the incompleteness
of the basis. We call this correction the \emph{Pulay} term in analogy
to a corresponding term -- the Pulay force -- in atomic-force calculations.\cite{LAPW-Forces-I,LAPW-Forces-II,footnote-II}
As already discussed, the first term is inaccurate because of the
truncation of the sum $\left(n'\le N\right)$. This inaccuracy is
corrected by the second term that arises from the explicit variation
of the basis due to a change in the effective potential. We call this
term the \emph{basis-response} (BR) correction. In the limit of a
complete basis (with an infinite number of states) the Pulay and BR
term would vanish, and the standard perturbation theory (SPT) expression
would give the exact result. The sum over states in the BR part can
be interpreted as a double-counting correction that subtracts response
contributions already contained in the first term.

So far, we have restricted the derivation to the augmented plane waves
defined in Eq.~\eqref{eq: APWs}. Corresponding corrections can be
derived for the local orbitals and the core states. While the derivation
for the former closely follows the steps already presented, the core
states require some different considerations. In contrast to the basis
functions, they fulfill the boundary condition that they approach
zero for $r\rightarrow\infty$. This boundary condition is used to
determine the energies of the core levels, which in contrast to the
construction of the LAPW basis functions are not chosen as parameters
but result from an atomic eigenvalue problem. Therefore, we use a
finite-difference approach: we solve the atomic eigenvalue problems
for the perturbed potentials $V_{\mathrm{eff},0}(r)+\frac{\lambda}{2}M_{I}(r)$
and $V_{\mathrm{eff},0}(r)-\frac{\lambda}{2}M_{I}(r)$ with $\lambda=0.0001$,
take the difference of the resulting core wave functions, and divide
by $\lambda$, which directly yields the linear response of the core
state. As the fully relativistic Dirac equation is employed for the
core states, the finite-difference approach yields the solution of
the fully relativistic Sternheimer equation.

Finally we use Eq.~\eqref{eq: wave functions response inkl FBC}
to construct the density response matrix {[}Eq.~\eqref{eq: response matrix}{]}
and the right-hand side of the OEP equation {[}Eq.~\eqref{eq: t vector}{]};
also consider Eqs.~\eqref{eq: density response} and \eqref{eq:phi_prime}.
As a result,

\begin{multline}
\chi_{\mathrm{s},IJ}=4\sum_{n\vec{k}}^{\mathrm{occ.}}\sum_{n'\le N}^{\mathrm{unocc.}}\left[\frac{\langle M_{I}\varphi_{n\vec{k}}|\varphi_{n'\vec{k}}\rangle\langle\varphi_{n'\vec{k}}|\varphi_{n\vec{k}}M_{J}\rangle}{\epsilon_{n\vec{k}}-\epsilon_{n'\vec{k}}}\right.\\
\left.\hphantom{2\sum_{n\vec{k}}^{\mathrm{occ.}}\sum_{n'\le N}^{\mathrm{unocc.}}}+\frac{\langle M_{I}\varphi_{n\vec{k}}|\varphi_{n'\vec{k}}\rangle\langle\varphi_{n'\vec{k}}|H-\epsilon_{n'\vec{k}}|\tilde{\varphi}_{n\vec{k},J}^{(1)}\rangle}{\epsilon_{n\vec{k}}-\epsilon_{n'\vec{k}}}\right]\\
+4\sum_{n\vec{k}}^{\mathrm{occ.}}\left[\langle M_{I}\varphi_{n\vec{k}}|\tilde{\varphi}_{n\vec{k},J}^{(1)}\rangle\hspace{2.5cm}\vphantom{\sum_{n'\le N}}\right.\\
\left.-\sum_{n'\le N}\langle M_{I}\varphi_{n\vec{k}}|\varphi_{n'\vec{k}}\rangle\langle\varphi_{n'\vec{k}}|\tilde{\varphi}_{n\vec{k},J}^{(1)}\rangle\right]\,,\label{eq: response matrix IBS}\end{multline}
 and\begin{multline}
t_{I}=4\sum_{n\vec{k}}^{\mathrm{occ.}}\sum_{n'\le N}^{\mathrm{unocc.}}\left[\frac{\langle M_{I}\varphi_{n\vec{k}}|\varphi_{n'\vec{k}}\rangle}{\epsilon_{n\vec{k}}-\epsilon_{n'\vec{k}}}\langle\varphi_{n'\vec{k}}|V_{\mathrm{x}}^{\mathrm{NL}}|\varphi_{n\vec{k}}\rangle\right.\\
\left.+\frac{\langle\tilde{\varphi}_{n\vec{k},I}^{(1)}|H-\epsilon_{n'\vec{k}}|\varphi_{n'\vec{k}}\rangle}{\epsilon_{n\vec{k}}-\epsilon_{n'\vec{k}}}\langle\varphi_{n'\vec{k}}|V_{\mathrm{x}}^{\mathrm{NL}}|\varphi_{n\vec{k}}\rangle\right]\\
\hspace{-4cm}+4\sum_{n\vec{k}}^{\mathrm{occ.}}\left[\langle\tilde{\varphi}_{n\vec{k},I}^{(1)}|V_{\mathrm{x}}^{\mathrm{NL}}|\varphi_{n\vec{k}}\rangle\vphantom{\sum_{n'\le N}}\right.\\
\left.\hphantom{+2\sum_{n\vec{k}}}-\sum_{n'\le N}\langle\tilde{\varphi}_{n\vec{k},I}^{(1)}|\varphi_{n'\vec{k}}\rangle\langle\varphi_{n'\vec{k}}|V_{\mathrm{x}}^{\mathrm{NL}}|\varphi_{n\vec{k}}\rangle\right]\,.\label{eq: t vector IBS-1}\end{multline}
 are again given as a sum over an SPT, Pulay, and BR term.

In its present form, the expression in Eq.~\eqref{eq: response matrix IBS}
breaks the Hermiticity of $\chi_{\mathrm{s},IJ}$ because the additional
term is formally asymmetric in the indices $I$ and $J$. However,
the numerical deviation from Hermiticity is small. To eliminate these
inaccuracies, we take the average $(\chi_{\mathrm{s},IJ}+\chi_{\mathrm{s},JI}^{*})/2$.

\section{Performance of the IBC\label{sec:Performance}}

We have implemented the incomplete-basis-set correction (IBC), as
described in the previous section, in the \noun{Fleur} program package,\cite{Fleur}
which is based on the FLAPW method. Before showing results for the
nitrides BN, AlN, GaN, InN, and ScN as well as the perovskites $\mathrm{CaTiO}_{3}$,
$\mathrm{SrTiO}_{3}$, and $\mathrm{BaTiO}_{3}$ in the next section,
we first analyze in detail how the convergence properties of the single-particle
response function {[}Eq.~\eqref{eq: response matrix IBS}{]}, the
local exchange potential, and the band gap are improved by the IBC
for the example of rock-salt scandium nitride. The improvements are
twofold: (1) the spherical response function converges with much smaller
LAPW basis sets than before; (2) for a given LAPW basis much fewer
unoccupied states are needed for its construction.

Unless noted otherwise, the LAPW cutoff parameters $G_{\mathrm{max}}=3.8\, a_{0}^{-1}$
and $l_{\mathrm{max}}=8$ were used for the calculations of rock-salt
$\mathrm{ScN}$ at the experimental lattice constant of $8.50\, a_{0}$
($a_{0}$ is the Bohr radius). The Sc $1s$, $2s$, and $2p$ states
as well as the N $1s$ state are treated as core states. All other
states -- including the $3s$ and $3p$ semicore states of Sc -- are
treated as valence. The Brillouin zone is sampled with a 4$\times$4$\times$4
$\vec{k}$-point mesh.

\subsection{Response function}

We remind the reader that the IBC consists of two terms, the Pulay
and the BR term, which are derived as a correction for the expression
of standard perturbation theory abbreviated by SPT. In principle,
the core states are included in Eq.~\eqref{eq: response matrix IBS}
as part of the sum over the occupied states. This is reminiscent of
the fact that the IBC not only corrects for the incompleteness of
the basis, but also comprises the exact core-state response, which
will give a numerically important contribution, but only a nearly
rigid shift of the convergence curves. To simplify the discussion,
we will leave the contribution of the core states out until later.%
\begin{figure}
\includegraphics[scale=0.33,angle=-90]{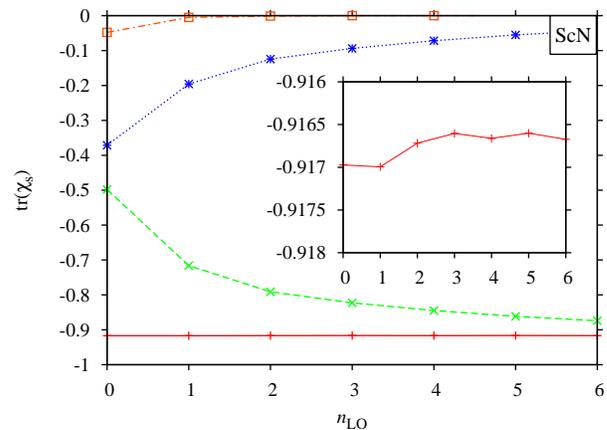}

\caption{(Color Online) Convergence of the trace of $\chi_{\mathrm{s}}$ for
$\mathrm{ScN}$ as a function of the number of local orbitals $n_{\mathrm{LO}}$
per $l$ quantum number $\left(0\le l\le4\right)$ and atom. The SPT
{[}(green) dashed{]}, BR {[}(blue) dotted{]}, and Pulay term {[}(orange)
dot-dashed{]} as well as their sum {[}(red) solid line{]} are shown,
respectively. We only consider spherical MPB functions, and the (constant)
contribution of core states is neglected. \label{fig: trace n-response}}

\end{figure}

Figure \ref{fig: trace n-response} shows the convergence of the trace
of the matrix $\chi_{\mathrm{s},IJ}$ {[}Eq.~\eqref{eq: response matrix IBS}{]}
as well as its SPT, Pulay, and BR contributions for ScN as a function
of the number of local orbitals $n_{\mathrm{LO}}$ added to the LAPW
basis for each $lm$ channel with $0\le l\le4$ and $|m|\le l$ (in
addition to local orbitals already used for the semicore $3s$ and
$3p$ states of scandium). The added local orbitals are placed at
energies in the conduction band according to the recipe of Ref.~\onlinecite{FLAPW-EXX}.
For simplicity, we have employed the same number of additional local
orbitals for Sc and N. In this case the MPB consists of 13 spherical
functions, seven at the Sc atom and six at the N atom. The trace $\mathrm{tr}(\chi_{\mathrm{s}})=\sum_{I}\chi_{\mathrm{s},II}$
is restricted to these functions. The slow convergence of the SPT
{[}dashed (green) line{]} is a direct consequence of the low flexibility
of the LAPW basis in the MT spheres with respect to changes of the
effective potential. The dotted (blue) and dot-dashed (orange) line
show the corresponding behavior of the BR and Pulay-term, respectively.
As expected, both corrections become smaller as the basis set becomes
more and more complete toward $n_{\mathrm{LO}}=6$. The major part
of the correction originates from the BR term, while the Pulay term
is significant only for $n_{\mathrm{LO}}=0$ and, even there, accounts
for merely 10\% of the total IBC. For $n_{\mathrm{LO}}\ge1$, it rapidly
approaches zero. The BR term is much more important numerically. In
fact, the dotted (blue) and dashed (green) lines appear to be mirror
images of one another, showing that the BR correction compensates
nearly exactly for what is missing in the SPT term. The sum of all
terms produces the solid (red) line, which appears to be constant
on the scale of the diagram. From the inset, which shows the curve
on a much finer scale, we see that the variations are below 0.05~\%,
an accuracy that we could never hope to achieve without the IBC. For
this particular case, we thus do not have to employ additional local
orbitals (i.e., $n_{\mathrm{LO}}=0$) for the unoccupied states.%
\begin{figure}
\includegraphics[scale=0.33,angle=-90]{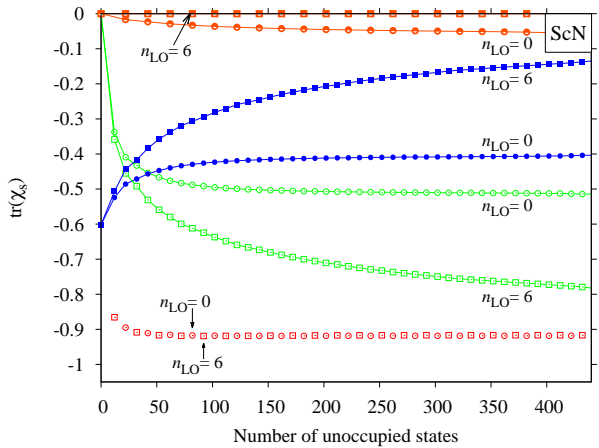}

\caption{(Color Online) Convergence of the trace of $\chi_{\mathrm{s}}$ for
$\mathrm{ScN}$ as a function of the number of unoccupied states $N-N_{\mathrm{el}}/2$
. The SPT {[}(green) open symbols{]}, BR {[}(blue) filled symbols{]},
and Pulay term {[}(orange) hall-filled symbols{]} as well as their
sum {[}(red) open symbols{]} are shown, respectively. Circles and
squares are used to distinguish the cases $n_{\mathrm{LO}}=0$ and
$n_{\mathrm{LO}}=6$. \label{fig:Convergence wrt bands}}

\end{figure}

Up to now, we have discussed how the IBC affects the convergence of
the single-particle response function with respect to the quality
of the LAPW basis. A perhaps more obvious convergence parameter is
the number of states $N$ included in the evaluation of Eq.~\eqref{eq: response matrix IBS}.
All three terms, SPT, Pulay, and BR, involve summations over the unoccupied
states up to the maximal band index $N$. Keeping in mind that the
sum in the BR part can be interpreted as a double-counting correction,
one can hope that the function $\tilde{\varphi}_{n\vec{k},I}^{(1)}(\vec{r})$
obtained nonperturbatively from direct integration of the Sternheimer
equation already contains, to a certain degree, information about
the whole infinite spectrum of unoccupied states. As a matter of fact,
this conjecture is substantiated by Fig.~\ref{fig:Convergence wrt bands},
which shows the convergence of the trace of $\chi_{\mathrm{s}}$ for
two different LAPW basis sets with $n_{\mathrm{LO}}=0$ (circles)
and $n_{\mathrm{LO}}=6$ (squares) as a function of the number of
unoccupied states $N-N_{\mathrm{el}}/2$, where $N_{\mathrm{el}}$
is the number of valence electrons per unit cell ($N_{\mathrm{el}}=16$
for ScN). (The reciprocal cutoff value of the LAPW basis has been
increased to $G_{\mathrm{max}}=5.5\, a_{0}^{-1}$ to generate up to
450 bands.) Similarly to Fig.~\ref{fig: trace n-response}, the SPT
term {[}(green) open symbols{]} shows a very slow convergence with
respect to the number of unoccupied states. We note that in the $n_{\mathrm{LO}}=0$
case {[}(green) open circles{]} a large part of the response is actually
missing in the SPT term resulting in a false convergence behavior
of the curve, which seems to converge, but toward a wrong value. As
above, the BR term {[}(blue) filled symbols{]} is much more important
than the Pulay term {[}(orange) half-filled circles{]} and counterbalances
the first term almost exactly. While the Pulay term is practically
zero in the case $n_{\mathrm{LO}}=6$, also compare Fig.~\ref{fig: trace n-response},
it is significant in the case $n_{\mathrm{LO}}=0$ and yields a small
but important contribution to the sum of all terms. In fact, this
sum nearly follows the same curve {[}(red) open symbols, circles and
squares correspond to $n_{\mathrm{LO}}=0$ and $n_{\mathrm{LO}}=6$,
respectively{]} in the two cases, showing again that with the IBC
we can restrict ourselves to the conventional LAPW basis, i.e., $n_{\mathrm{LO}}=0$.
The total sum converges extremely fast, thanks to the BR term, in
which the infinite spectrum of eigenstates is already incorporated
to a large extent. Without showing further results we note that the
IBC affects the convergence of the right hand side {[}Eq.~\eqref{eq: t vector IBS-1}{]}
in a similarly beneficial way.

So far, we have not discussed the contribution of the core states
to the density response $\chi_{\mathrm{s}}$. For the core state response
we solve a fully relativistic Sternheimer equation by a finite-difference
approach as discussed in Sec.~\ref{sec: FBSC}. The resulting solution
embodies the full infinite spectrum of unoccupied states by construction
and can, therefore, be considered to represent already the exact core-state
response (up to numerical errors connected with the finite-difference
approach, which can be made arbitrarily small, though). Therefore,
we can set $N$ to the number of occupied states in Eq.~\eqref{eq: response matrix IBS}.
The first term is then zero, and only the BR term remains.

The contribution of the core states to the response function is numerically
important. In the case of ScN it is more than four times larger than
the valence contribution. However, it should be noted that the effective
quantity for the construction of the local exchange potential is not
the response function itself, but its inverse {[}see Eq.~\eqref{eq: OEP equation}{]}.
Therefore, large eigenvalues of $\chi_{\mathrm{s}}$ become comparatively
unimportant in $\chi_{\mathrm{s}}^{-1}$. This is confirmed by the
following observation. We find that the SPT expression alone is incapable
of describing the core-state response. Even with $n_{\mathrm{LO}}=6$
and $N=450$, only about $20\%$ of the contribution of the core states
is accounted for, which manifests the hardly surprising fact that
the LAPW valence basis is unsuitable to describe changes in the core
states. However, this shortcoming affects the resulting exchange potential
and KS transition energies only slightly, as we will see below.

\subsection{Local exchange potential}

Figure \ref{fig: ScN-100} shows the local exact exchange potential
between neighboring scandium and nitrogen atoms for three calculations,
all of them without any local orbitals for the unoccupied states.
For these calculations an MPB with $G_{\mathrm{max}}^{\prime}=2.8\, a_{0}^{-1}$
and $L_{\mathrm{max}}=4$ is employed, and the IBC is applied to both
core and valence electrons. Without the IBC (green dashed line) we
obtain an unphysical, strongly varying potential that even tends to
an unreasonable positive value at the position of the nitrogen nucleus.
In Ref.~\onlinecite{FLAPW-EXX} we showed that the spurious oscillations
can be avoided by augmenting the LAPW basis with local orbitals leading
to a smooth and physical potential, at the expense of a very costly
calculation. As shown in Fig.~\ref{fig: ScN-100}, the same smooth
shape of the local exact exchange potential is realized by employing
the IBC with a considerably smaller computational overhead. We emphasize
that no local orbitals are used in this calculation.

When one watches closely, one still sees very slight anomalies of
the dotted curve at the atomic nuclei, here more pronounced at the
nitrogen nucleus (also cf.~the insets). These result from the fact
that the radial MT potential enters with a factor $r^{2}$ into the
equations. The region close to the nuclei has only little weight and
is therefore difficult to converge. On the one hand, for small $r$
the total effective potential is dominated by the potential of the
nucleus $-Z/r$ as well as the centrifugal kinetic energy barrier
$l(l+1)/(2r^{2})$ such that slight inaccuracies at the nuclei prove
to be irrelevant in the calculation of the electronic structure. On
the other hand, we can find a simple remedy by an additional constraint
for the spherical functions $(L=0)$ of the MPB, which has the additional
benefit of reducing their number by one for each atom in the unit
cell. We require the gradient of these functions to vanish at $r=0$.
As already mentioned in Sec.~\ref{sec: FLAPW-Method}, this behavior
has been observed for the local exact exchange potential in previous
publications.\cite{Zero-Slope-1,Zero-Slope-2,Zero-Slope-3} Thus,
the constraint does not induce errors. With such a modified MPB the
anomalies of $V_{\mathrm{x}}(\vec{r})$ at the nuclei disappear, and
we obtain the red solid line shown in Fig.~\ref{fig: ScN-100}. We
also note that the numerical stability benefits from this modification.%
\begin{figure}
\includegraphics[scale=0.33,angle=-90]{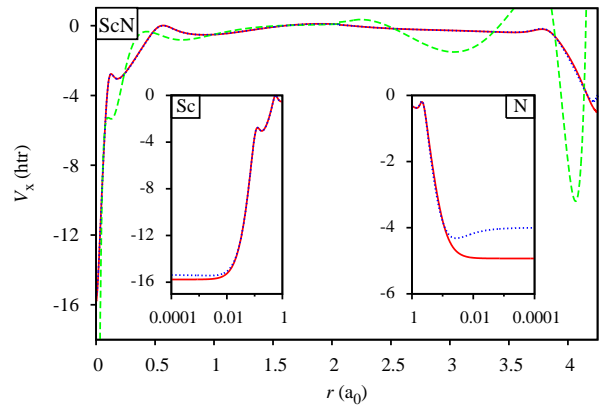}

\caption{(Color Online) Local EXX potential for ScN on a line connecting the
neighboring Sc and N atoms. We have used a conventional LAPW basis
without local orbitals for the unoccupied states. The (blue) dotted
and (green) dashed curves correspond to calculations with and without
the IBC. For the (red) solid curve we have employed an additional
constraint for the MPB (see text). Eq.~\eqref{eq: OEP equation}
defines the potential only up to a constant. Here, we use the convention
$\int V_{\mathrm{x}}(\vec{r})\, d^{3}r=0$.\label{fig: ScN-100}}

\end{figure}

\begin{figure}
(a)\includegraphics[scale=0.5,angle=-90]{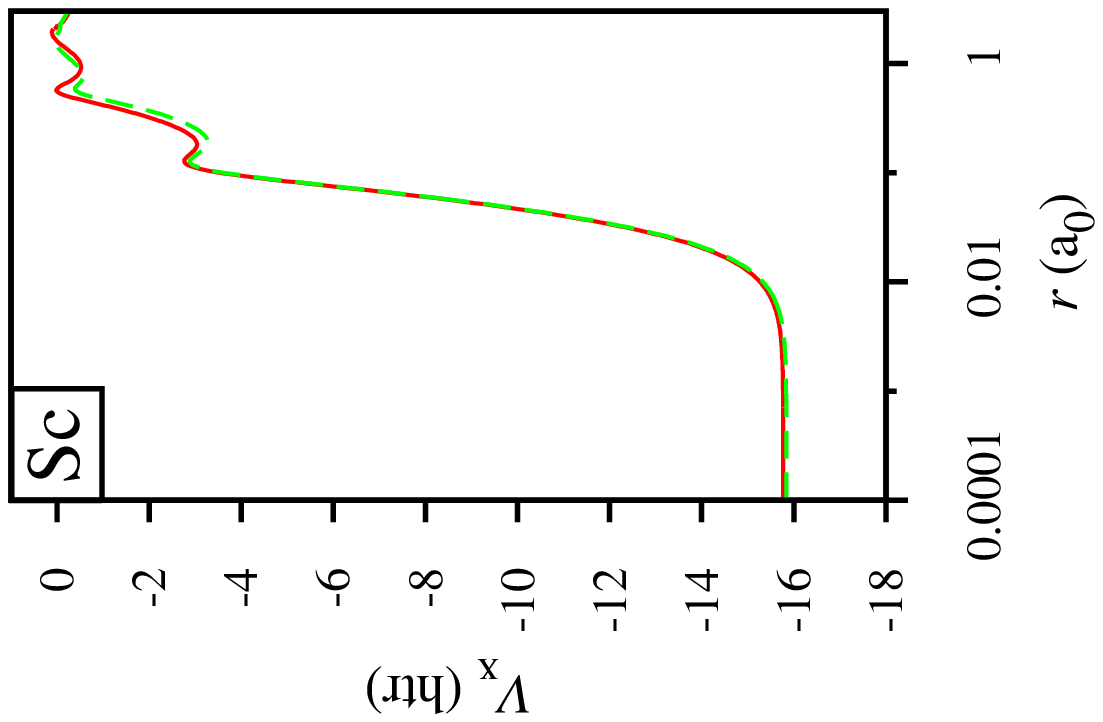}(b)\includegraphics[scale=0.5,angle=-90]{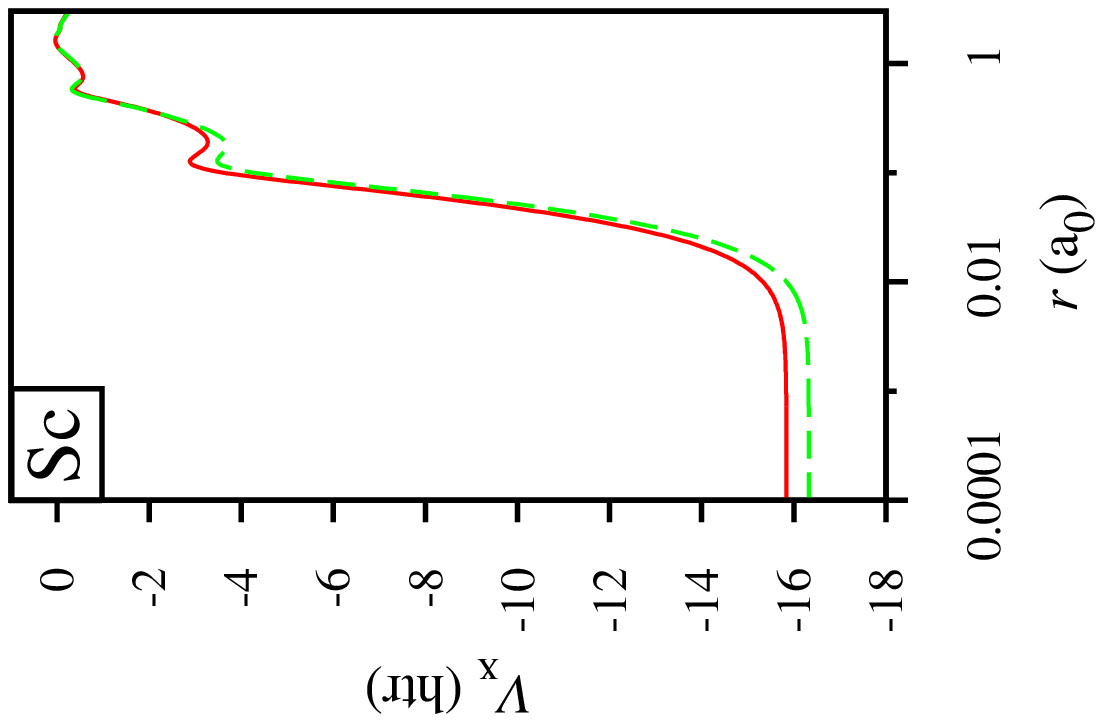}

\caption{(Color Online) (a) Comparison of $V_{\mathrm{x}}(\vec{r})$ close
to the atomic nucleus of Sc for calculations with {[}(green) dashed
line, $n_{\mathrm{LO}}=1${]} and without {[}(red) solid line, $n_{\mathrm{LO}}=0${]}
local orbitals. (b) $V_{\mathrm{x}}(\vec{r})$ with {[}(red) solid
line{]} and without {[}(green) dashed line{]} the core response of
the IBC.\label{fig: ScN lo/core}}

\end{figure}

As already pointed out, the IBC is presently only applied to spherical
variations of the potential. To converge the nonspherical contributions
properly we still need a few local orbitals. Figure \ref{fig: ScN lo/core}(a)
shows this effect, which is strongest close to the atomic nucleus
of Sc. Using a single set of local orbitals, i.e., $n_{\mathrm{LO}}=1$,
gives a slight correction of the potential, which gives rise to changes
in single-particle transition energies in the order of $0.10-0.15\,\mathrm{eV}$.
(We have considered the gap transitions $\Gamma\rightarrow\Gamma$,
$\Gamma\rightarrow\mathrm{X}$, and $\Gamma\rightarrow\mathrm{L}$.)
For $n_{\mathrm{LO}}=2$ changes are less than $0.03\,\mathrm{eV}$.
It is important to note that the calculations always converge to the
same result, irrespective of whether or not the IBC is used. We expect
that once the IBC is extended to the nonspherical MPB functions, no
extra local orbitals are required anymore.

As already discussed, taking into account the exact core response
yields a numerically important contribution to the single-particle
response function. To demonstrate the effect on the local exchange
potential, we switch the IBC for the core states off. Figure \ref{fig: ScN lo/core}(b)
shows that this has only a comparatively small effect on the shape
of the resulting potential. At first sight, the incurred changes in
the potential are more pronounced than in Fig.~\ref{fig: ScN lo/core}(a),
but, in contrast to Fig.~\ref{fig: ScN lo/core}(a), they mostly
affect the region close to the nucleus, where the effective potential
is dominated by the potential of the nucleus and the angular kinetic
energy. In fact, the single-particle transition energies are influenced
only little: they change by less than 0.04 eV. However, it should
be noted that the exact core response is evaluated at virtually no
extra cost.%

\subsection{KS band gap}

After discussing the effect of the IBC on the ingredients of the OEP
integral equation and the local exact exchange potential, we now turn
to the KS band gap. As we have presently derived the IBC only for
the spherical MPB functions, we employ the cutoff $L_{\mathrm{max}}=0$
for this test calculation in order to highlight the improvements resulting
from the IBC. This will produce a potential $V_{\mathrm{x}}(r)$ that
is purely spherical in the muffin-tin spheres. We note again that
a generalization to nonspherical functions $(L>0)$ is possible, but
not yet implemented. Furthermore, we restrict the interstitial exchange
potential to a constant for simplicity.

\begin{figure}
\includegraphics[scale=0.34,angle=-90]{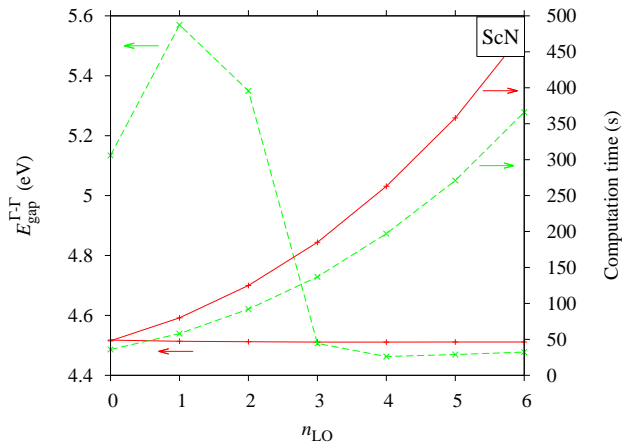}

\caption{(Color Online) Convergence of the direct gap of ScN with respect to
the number of local orbitals $(n_{\mathrm{LO}})$ added to the LAPW
basis for one-shot EXX calculations with {[}(red) solid line{]} and
without {[}(green) dashed line){]} the IBC. The computation time is
shown on the right scale. \label{fig: EXX-ASA gap+timing}}

\end{figure}

Figure~\ref{fig: EXX-ASA gap+timing} shows the direct $(\Gamma\rightarrow\Gamma)$
gap of ScN for this specific numerical setup as a function of the
number of local orbitals per $l$ channel added to the LAPW basis.
The case $n_{\mathrm{LO}}=6$ corresponds to $300$ additional basis
functions. It is known that small inaccuracies in the band gap can
occur due to the linearization error of the LAPW basis (in the case
$n_{\mathrm{LO}}=0$). To eliminate this pure basis-set effect, we
have taken two measures: (1) we have performed only a single EXX-OEP
iteration (starting from a PBE\cite{GGA-PBE} ground state), and (2)
the final diagonalization of the KS Hamiltonian was performed with
the most accurate basis set $(n_{\mathrm{LO}}=6)$. The value $n_{\mathrm{LO}}$
shown on the abscissa thus corresponds to the basis used in solving
the OEP equation, and variations in the band gap can be attributed
exclusively to the precision of the local exchange potential without
additional basis-set effects. We note that, in spite of the very small
MPB used here and in spite of performing only one iteration, the resulting
band gap is surprisingly close to the fully converged one (see~Tab.~\ref{tab: III-V nitrides}).

Without the IBC {[}(green) dashed curve{]} four sets of local orbitals
$(n_{\mathrm{LO}}=4)$ are necessary to obtain a direct gap with an
accuracy of $0.05\,\mathrm{eV}$ (we note that, judging from the form
of the curve, the accuracy at $n_{\mathrm{LO}}=3$ seems to be due
to a fortuitous cancellation of errors). On the other hand, the calculation
with the IBC yields a reliable gap with an accuracy of $0.007\,\mathrm{eV}$
already without any local orbitals for the unoccupied states $(n_{\mathrm{LO}}=0)$.
Both curves converge to the same band-gap value, while the convergence
of the IBC values is hardly visible on the scale of the diagram. We
have indicated the computation time on the right scale. The computational
overhead of the IBC calculations is due to the additional evaluation
of the Pulay and BR term, in particular for the matrix elements $\langle\varphi_{n\vec{k}}|V_{x}^{\mathrm{NL}}|\tilde{\varphi}_{n\vec{k},I}\rangle$
in Eq.~\eqref{eq: t vector IBS-1}. It is difficult to compare the
efficiency of the calculations, since the least accurate calculation
with the IBC is still more accurate than the most accurate one without.
If we take an accuracy of $0.05\,\mathrm{eV}$ as a criterion, one
would deduce an acceleration of the code by a factor of four. Using
less unoccupied states (cf.~Fig.~\ref{fig:Convergence wrt bands})
could further reduce the computation time.%

\section{Results and Discussion\label{sec: results}}

In Table \ref{tab: III-V nitrides} we report KS transition energies,
i.e., KS eigenvalue differences, for the III-V nitrides in the zincblende
structure and for rock-salt ScN, calculated with the EXX and EXXc
functionals. The latter contains additionally the LDA correlation
functional in the parametrization of Perdew and Zunger.\cite{LDA-PZ}
While the ground-state crystal structure of the III-V nitrides is
wurtzite, they can be synthesized in the metastable zincblende structure
by epitaxial growth techniques. All calculations were performed at
the experimental lattice constants (ScN, $8.50\, a_{0}$; BN, $6.84\, a_{0}$;
AlN, $8.26\, a_{0}$; GaN $8.50\, a_{0}$; InN $9.41\, a_{0}$) with
an 8$\times$8$\times$8 $\vec{k}$-point sampling and with local
orbitals for the complete semicore shell (e.g., $2s$ and $2p$ states
of Al). The numerical cutoff parameters $G_{\mathrm{max}}$, $G'_{\mathrm{max}}$,
$l_{\mathrm{max}}$, and $L_{\mathrm{max}}$ as well as the number
of local orbitals are determined such that the transition energies
are converged to within $10\,\mathrm{meV}$. We compare our results
with plane-wave pseudopotential calculations and experimental values
from the literature.

The EXX and EXXc functionals give KS transition energies that are
much closer to the experimental value than LDA. InN and ScN, which
are metallic in LDA, are correctly predicted to be semiconductors.
The inclusion of the LDA correlation functional in EXXc increases
the values by about $0.01-0.2\,\mathrm{eV}$, but does not yield a
systematic improvement when compared to experiment. For AlN and ScN
our EXXc values agree well with those from the pseudopotential studies.
The differences are mostly of the same order as the differences in
the LDA transition energies.

In the case of GaN and InN the situation is more complex because of
the semicore $d$ states. In pseudopotential calculations they can
be treated approximately as atomic core states in the pseudopotential
or, at a considerably larger computational expense, as valence electrons
within the plane-wave basis. Therefore, Table \ref{tab: III-V nitrides}
lists two values for each system and functional from calculations
where the $d$ semicore electrons are treated as valence (\textquotedbl{}with
$d$\textquotedbl{}) or not (\textquotedbl{}no $d$\textquotedbl{}).
The former values, which are systematically smaller than the latter,
must be considered to be more accurate. Of course, we do not have
to make such a distinction in our calculations, because the FLAPW
method treats valence and core electrons, down to the $1s$ state,
on an equal footing. Indeed, our LDA transition energies compare better
with the pseudopotential calculations \textquotedbl{}with $d$\textquotedbl{}.
However, we find a much larger discrepancy for the EXXc functional.
In all cases the all-electron values fall in-between the two pseudopotential
results, often being equally far away from either value. Engel and
Schmid\cite{EXX-Transition-Oxides} pointed out for the case of transition
metal monoxides that the complete semicore shell, comprising the $3s$,
$3p$, and $3d$ states, must be treated as valence to obtain accurate
EXX values within the pseudopotential approach. It is to be expected
that this also holds for GaN and InN, which would explain the discrepancies
observed for the pseudopotential calculations with respect to the
all-electron results.%
\begin{table*}
\caption{KS transition energies (in eV) obtained with the LDA, EXX, and EXXc
potentials for the III-V nitrides and $\mathrm{ScN}$. An $8\times8\times8$
$\vec{k}$-point sampling has been employed for all calculations.
For comparison, plane-wave pseudopotential results and experimental
values from the literature are given. \label{tab: III-V nitrides}}

\begin{ruledtabular}

\begin{tabular}{ccrrrccccc}
 &  & \multicolumn{3}{c}{This work} & \multicolumn{4}{c}{Plane-wave PP} & \tabularnewline
\hline 
 &  & LDA  & EXX  & EXXc  & \multicolumn{2}{c}{LDA} & \multicolumn{2}{c}{EXXc} & Expt.\tabularnewline
$\mathrm{ScN}$  & $\Gamma\rightarrow\Gamma$  & $2.35$  & $4.41$  & $4.42$  & \multicolumn{2}{c}{$\hphantom{-}2.34^{c}$} & \multicolumn{2}{c}{$4.59^{c}$, $4.7^{d}\hphantom{0}$} & $3.8^{d}$\tabularnewline
 & $\Gamma\rightarrow\mathrm{X}$  & $-0.14$  & $1.58$  & $1.67$  & \multicolumn{2}{c}{$-0.15^{c}$} & \multicolumn{2}{c}{$1.7^{c}\hphantom{0}$,$1.6^{d}\hphantom{0}$} & $1.3^{d}$\tabularnewline
 & $\mathrm{X}\rightarrow\mathrm{X}$  & $0.79$  & $2.48$  & $2.56$  & \multicolumn{2}{c}{$\hphantom{-}0.75^{c}$} & \multicolumn{2}{c}{$2.59^{c}$,$2.9^{d}\hphantom{0}$} & $2.40^{d}$\tabularnewline
$\mathrm{BN}$  & $\Gamma\rightarrow\Gamma$  & $8.68$  & $9.80$  & $9.88$  &  &  & \multicolumn{2}{c}{} & \tabularnewline
 & $\Gamma\rightarrow\mathrm{L}$  & $10.18$  & $10.88$  & $10.95$  &  &  & \multicolumn{2}{c}{} & \tabularnewline
 & $\Gamma\rightarrow\mathrm{X}$  & $4.34$  & $5.42$  & $5.58$  &  &  & \multicolumn{2}{c}{} & $6.4^{e}$\tabularnewline
$\mathrm{AlN}$  & $\Gamma\rightarrow\Gamma$  & $4.20$  & $5.46$  & $5.59$  & \multicolumn{2}{c}{$\hphantom{-}4.27^{b}$} & \multicolumn{2}{c}{$5.66^{a}$,$5.74^{b}$} & $5.93^{f}$\tabularnewline
 & $\Gamma\rightarrow\mathrm{L}$  & $7.24$  & $8.42$  & $8.55$  & \multicolumn{2}{c}{$\hphantom{-}7.25^{b}$} & \multicolumn{2}{c}{$8.58^{b}$} & \tabularnewline
 & $\Gamma\rightarrow\mathrm{X}$  & $3.22$  & $4.77$  & $4.96$  & \multicolumn{2}{c}{$\hphantom{-}3.27^{b}$} & \multicolumn{2}{c}{$5.03^{a}$,$5.06^{b}$} & $5.3^{f}$\tabularnewline
 &  &  &  &  &  &  &  &  & \tabularnewline
 &  &  &  &  & \underbar{no $d$}  & \underbar{with $d$}  & \underbar{no $d$}  & \underbar{with $d$}  & \tabularnewline
$\mathrm{GaN}$  & $\Gamma\rightarrow\Gamma$  & $1.76$  & $3.11$  & $3.24$  & $2.22^{b}$  & $\hphantom{-}1.65^{b}$  & $3.52^{b}$  & $2.88^{b}$  & $3.27^{g}$\tabularnewline
 & $\Gamma\rightarrow\mathrm{L}$  & $4.55$  & $5.94$  & $6.05$  & $4.88^{b}$  & $\hphantom{-}4.43^{b}$  & $6.23^{b}$  & $5.64^{b}$  & \tabularnewline
 & $\Gamma\rightarrow\mathrm{X}$  & $3.25$  & $4.61$  & $4.80$  & $3.43^{b}$  & $\hphantom{-}3.30^{b}$  & $4.99^{b}$  & $4.66^{b}$  & \tabularnewline
$\mathrm{InN}$  & $\Gamma\rightarrow\Gamma$  & $-0.41$  & $0.98$  & $1.12$  & $0.27^{b}$  & $-0.44^{b}$  & $1.49^{b}$  & $0.81^{b}$  & $0.596^{h}$,$0.61^{g}$\tabularnewline
 & $\Gamma\rightarrow\mathrm{L}$  & $3.01$  & $4.37$  & $4.50$  & $3.51^{b}$  & $\hphantom{-}2.95^{b}$  & $4.75^{b}$  & $4.14^{b}$  & \tabularnewline
 & $\Gamma\rightarrow\mathrm{X}$  & $2.83$  & $4.23$  & $4.42$  & $2.87^{b}$  & $\hphantom{-}2.82^{b}$  & $4.63^{b}$  & $4.20^{b}$  & \tabularnewline
\end{tabular}

\end{ruledtabular}

\begin{raggedright}\begin{tabular}{ccccc}
$^{a}$Reference  & \onlinecite{EXX-PP-Staedele-2}  &  &  & \tabularnewline
$^{b}$Reference  & \onlinecite{EXX-Nitrides-Qteish}  &  &  & \tabularnewline
$^{c}$Reference  & \onlinecite{EXX-ScN-1}  &  &  & \tabularnewline
$^{d}$Reference  & \onlinecite{EXX-ScN-2}  &  &  & \tabularnewline
$^{e}$Reference  & \onlinecite{BN-cubic}  &  &  & \tabularnewline
$^{f}$Reference  & \onlinecite{AlN-cubic}  &  &  & \tabularnewline
$^{g}$Reference  & \onlinecite{GaN-cubic}  &  &  & \tabularnewline
$^{h}$Reference  & \onlinecite{InN-cubic-I}  &  &  & \tabularnewline
$^{g}$Reference  & \onlinecite{InN-cubic-II}  &  &  & \tabularnewline
\end{tabular}

\end{raggedright} 
\end{table*}

Table \ref{tab: d-band position} shows the energetic positions of
the $d$ levels in GaN and InN with respect to the Fermi level, which
is fixed at the valence band edge. It is well known that the LDA underbinds
the $3d$ states, which is usually attributed to the self-interaction
error of LDA. However, in spite of the fact that the EXX(c) is free
of this error, the position of the $d$ bands is hardly improved.
While in GaAs the $d$ levels are lowered by about $1\,\mathrm{eV}$,
they remain nearly at the same energy in InN. In comparison to the
experimental values, $17.7\,\mathrm{eV}$ (GaN) and $14.9\,\mathrm{eV}$
(InN; measured for wurtzite structure), this is hardly an improvement
over LDA. On the other hand, we find that the HF method yields $d$-band
positions at $21.28\,\mathrm{eV}$ (GaN) and $18.22\,\mathrm{eV}$
(InN), respectively, i.e., even below the experiment. As the HF and
EXX-OEP methods employ the same energy functional, the question arises
why the $d$-band positions appear at different energies. In short,
what is the meaning of the EXX-OEP single-particle energies? According
to Koopmans' theorem,\cite{Koopman} the single-particle energies
in the HF approach can be understood as total-energy differences between
an excited state and the ground state, neglecting orbital relaxation
effects. For the $d$ states, $\Delta E=E_{d}^{N-1}-E_{0}^{N}=\langle d|h^{\mathrm{HF}}|d\rangle=\epsilon_{d}^{\mathrm{HF}}$,
where the two total energies correspond to the many-body states with
and without a hole in a $d$ level, $\epsilon_{d}^{\mathrm{HF}}$
is the HF single-particle energy of that $d$ level, and $h^{\mathrm{HF}}$
is the HF single-particle Hamiltonian. The OEP approach was originally
intended as a procedure to simplify the solution of the HF equations
by introducing a local potential that minimizes the HF total energy.
In fact, the EXX-OEP and HF ground-state total energies are nearly
identical.\cite{FLAPW-EXX,Balance-Gaussian} This indicates that the
slight difference in shape of the single-particle wave functions only
has a negligible effect on the ground-state total energy $E_{0}^{N}$,
a statement that should also hold for the hole-state energy $E_{d}^{N-1}$.
Using Koopman's theorem again with the EXX-OEP wave functions instead
of the HF ones, we should obtain practically the same $\Delta E$.
However, as seen in Table~\ref{tab: d-band position}, the HF eigenvalue
$\epsilon_{d}^{\mathrm{HF}}$ differs from the corresponding EXX-OEP
eigenvalue $\epsilon_{d}^{\mathrm{EXX}}$ by $\Delta=\epsilon_{d}^{\mathrm{HF}}-\epsilon_{d}^{\mathrm{EXX}}=\langle d|V_{\mathrm{x}}^{\mathrm{NL}}-V_{\mathrm{x}}|d\rangle$.
As a consequence, the underbinding of the $d$ bands in LDA -- and
also in EXX(c) -- cannot be solely ascribed to the self-interaction
error contrary to common belief. Part of this underbinding is due
to the \textit{forced locality} of the effective potential, which
gives rise to a nonzero energy shift that is very similar in spirit
to the derivative discontinuity\cite{KS-GAP1,KS-GAP2} of the xc functional.
Even with the exact functional we cannot expect that the $d$-band
positions lie at the experimental positions in the same way as we
cannot expect the KS band gap to equal the experimental gap.%
\begin{table}
\caption{Energetic positions of the $d$ levels in GaN and InN in eV with respect
to the Fermi energy and averaged over the Brillouin zone and the $d$
bands.\label{tab: d-band position}}

\begin{ruledtabular}

\begin{tabular}{cccccc}
 & LDA  & EXX  & EXXc  & HF  & Expt.\tabularnewline
$\mathrm{GaN}$  & $-13.08$  & $-13.93$  & $-14.15$  & $-21.28$  & $-17.7^{a}$\tabularnewline
$\mathrm{InN}$  & $-12.72$  & $-12.65$  & $-12.83$  & $-18.22$  & $-14.9^{b}$\tabularnewline
\end{tabular}

\end{ruledtabular}

\begin{raggedright}\begin{tabular}{cl}
$^{a}$Reference  & \onlinecite{GaN-cubic-d-position}\tabularnewline
$^{b}$Reference  & \onlinecite{InN-wz-I} (Measured for the wurtzite compound.)\tabularnewline
\end{tabular}

\end{raggedright} 
\end{table}

\begin{figure*}
\includegraphics[scale=0.55,angle=-90]{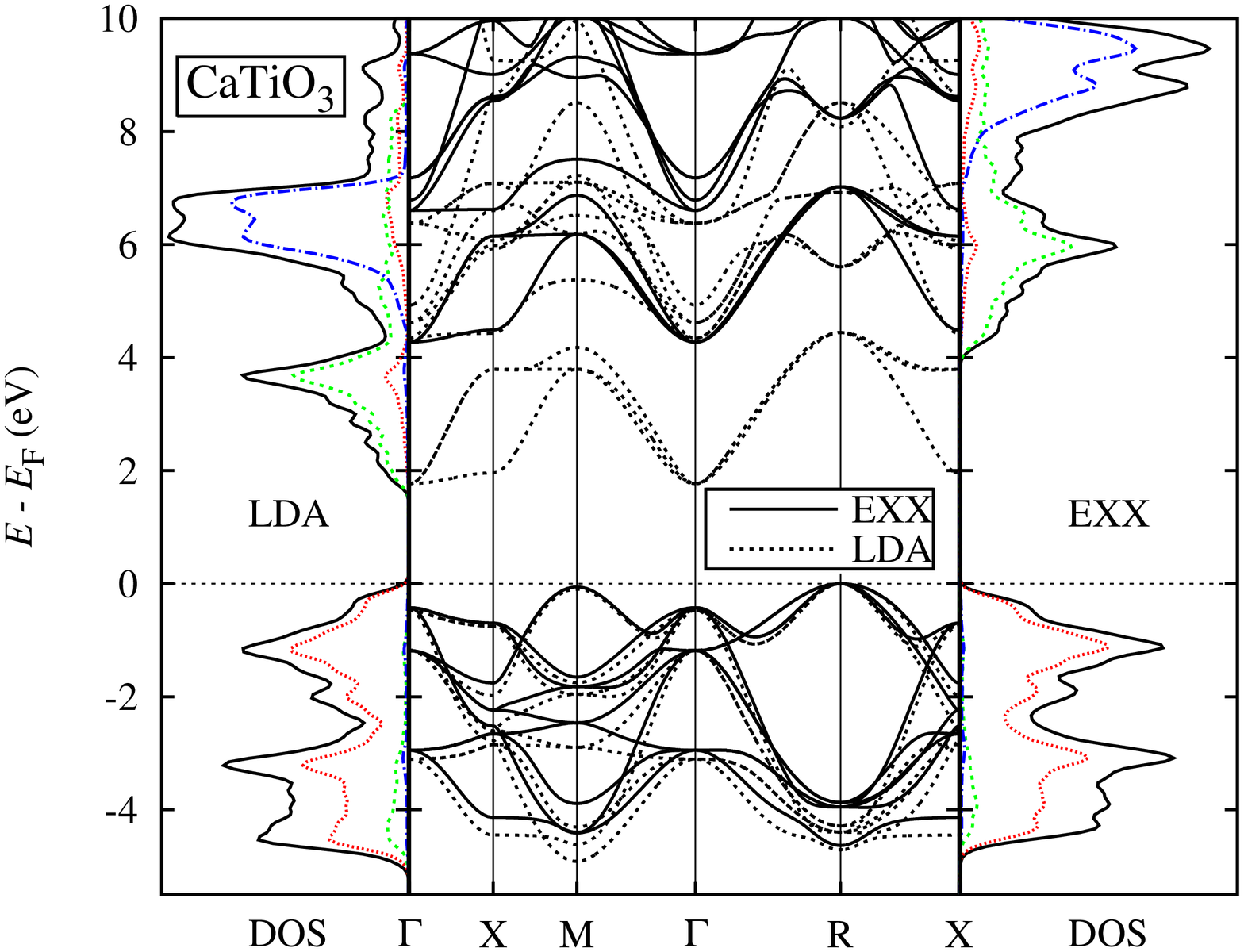}

\caption{(Color Online) Comparison of the LDA and EXX band structures and densities
of states (DOS) for $\mathrm{CaTiO_{3}}$. The total and partial DOS
for Ca $3d$, Ti $3d$, and O $2p$ are shown as (black) solid, (blue)
dot-dashed, (green) dashed, and (red) dotted lines, respectively.\label{fig: DOS/bandstructure CaTiO3}}

\end{figure*}

Finally, we report results for the electronic structure of the perovskite
transition-metal oxides $\mathrm{CaTiO_{3}}$, $\mathrm{SrTiO_{3}}$
and $\mathrm{BaTiO_{3}}$. We have calculated the materials in the
ideal cubic structure at experimental lattice constants ($\mathrm{CaTiO}_{3},\,7.35\, a_{\mathrm{0}}$;
$\mathrm{SrTiO}_{3},\,7.46\, a_{\mathrm{0}}$; $\mathrm{BaTiO_{3},\,7.60\, a_{\mathrm{0}}}$)
and with the functionals LDA, EXX, and EXXc. The Ti $3s$ and $3p$
states as well as the first subshell of the cations (Ca, $3s$~$3p$;
Sr, $4s$~$4p$; Ba $5s$~$5p$) have been treated with local orbitals.
The electronic structures of the three systems are very similar. They
are semiconductors with an indirect band gap between the $\mathrm{R}$
and the $\Gamma$ point and a direct gap at the $\Gamma$ point. We
compare the LDA and EXX band structures and densities of states of
$\mathrm{CaTiO_{3}}$ in Fig.~\ref{fig: DOS/bandstructure CaTiO3}.
The EXX functional shifts the Ti $3d$ conduction bands away from
the O $2p$ valence bands and thus opens the band gap, while the band
dispersions remain nearly unaffected. The valence band width is slightly
decreased, though.%

KS transition energies are reported in Table \ref{tab: perovskites}.
The LDA underestimates the energies by roughly $50\%$. Although the
EXX(c) functionals yield values that are closer to experiment, we
observe a pronounced overestimation by about $25\%$. This might be
due to two reasons: (1) The fortuitous error cancellation arising
from the neglect of correlation on the one hand and of the derivative
discontinuity on the other hand does not work for these transition-metal
oxides. (2) The ideal cubic structure used in our study does not correspond
to the experimentally measured systems. It is known that these perovskite
materials undergo a number of structural phase transitions. For example,
$\mathrm{SrTiO_{3}}$ is cubic at room temperature but tetragonal
at low temperatures. $\mathrm{BaTiO_{3}}$, on the contrary, is cubic
at elevated temperatures and tetragonal at room temperature.

\begin{table}
\caption{Direct and indirect KS band gaps (in eV) for cubic $\mathrm{CaTiO}_{3}$,
$\mathrm{SrTiO}_{3}$, and $\mathrm{BaTiO}_{3}$ compared with experimental
values. A 6$\times$6$\times$6 $\vec{k}$-point sampling was employed.\label{tab: perovskites}}

\begin{ruledtabular}

\begin{tabular}{cccccc}
 &  & LDA  & EXX  & EXXc  & Expt.\tabularnewline
$\mathrm{CaTiO}_{3}$  & $\Gamma\rightarrow\Gamma$  & $2.24$  & $4.70$  & $4.74$  & \tabularnewline
 & $\mathrm{R}\rightarrow\Gamma$  & $1.77$  & $4.28$  & $4.31$  & $3.57^{a}$\tabularnewline
 &  &  &  &  & \tabularnewline
$\mathrm{SrTiO}_{3}$  & $\Gamma\rightarrow\Gamma$  & $2.09$  & $4.51$  & $4.54$  & $3.75^{b}$\tabularnewline
 & $\mathrm{R}\rightarrow\Gamma$  & $1.73$  & $4.20$  & $4.22$  & $3.25^{b}$\tabularnewline
 &  &  &  &  & \tabularnewline
$\mathrm{BaTiO}_{3}$  & $\Gamma\rightarrow\Gamma$  & $1.81$  & $4.08$  & $4.12$  & \tabularnewline
 & $\mathrm{R}\rightarrow\Gamma$  & $1.7$  & $4.08$  & $4.11$  & $3.2^{c}$\tabularnewline
\end{tabular}

\end{ruledtabular}

\begin{raggedright}\begin{tabular}{ccccc}
$^{a}$Reference  & \onlinecite{CaTiO3-exp}  &  &  & \tabularnewline
$^{b}$Reference  & \onlinecite{SrTiO3-exp}  &  &  & \tabularnewline
$^{c}$Reference  & \onlinecite{BaTiO3-exp}  &  &  & \tabularnewline
\end{tabular}

\end{raggedright} 
\end{table}

\section{Conclusions\label{sec: Conclusions}}

We have described an efficient way to calculate precise all-electron
response functions within the FLAPW method. The key is the development
of an incomplete-basis-set correction (IBC), which was derived by
describing the response of the LAPW basis functions to changes of
the effective potential explicitly by means of radial Sternheimer
equations. In this way, the IBC incorporates important response contributions
that lie outside the Hilbert space formed by the LAPW basis set. The
resulting formula for the response function consists of three terms:
the conventional sum-over-states expression of standard perturbation
theory (SPT term) and two additional terms, the \textit{basis-response}
and the \textit{Pulay} term, whose mathematical expression resembles
that of the Pulay force of atomic-force calculations. The basis-response
term is more important numerically than the Pulay term. Both vanish
in the limit of a complete, infinite basis. Together they constitute
the IBC, which also yields an explicit and, in principle, exact treatment
of the response of the core states that avoids the sum-over-states
expression. The total correction is not small. It can be as large
as the SPT term.

As a practical example, we have employed the IBC to the EXX-OEP method,
an approach that allows the construction of a local exchange potential
from the orbital-dependent EXX functional. It involves two response
quantities: the density response function and a response function
for the single-particle states. We have implemented the IBC for both
and demonstrated explicitly for the case of rock-salt scandium nitride
that it improves the convergence with respect to the LAPW basis and
the number of unoccupied states considerably. While without the correction
the solution of the OEP equation requires a highly converged LAPW
basis with a large number of local orbitals,\cite{FLAPW-EXX} no extra
local orbitals are needed when we use the IBC. A similar statement
can be made about the number of unoccupied states in the sum-over-states
expression related to perturbation theory. As the numerical solution
of the Sternheimer equation already incorporates the infinite single-particle
spectrum to a certain degree, the response function converges at much
fewer bands.

With this new scheme we have performed all-electron EXX(c)-OEP calculations
for the III-V nitrides in the zincblende structure as well as rock-salt
ScN. Our results agree favorably with plane-wave pseudopotential EXXc
calculations from the literature. However, larger deviations are observed
for GaN and InN, which we attributed to the neglect of semicore states
($3s$, $3p$ of Ga and $4s$, $4p$ of In) in the pseudopotential
calculations. This is in accordance with a recent publication by Engel
and Schmid.\cite{EXX-Transition-Oxides} Despite the fact that the
EXX functional is self-interaction-free, it does not necessarily improve
the position of the $d$ states with respect to LDA, as shown for
the semicore $d$ states of GaN and InN. We have explained this observation
by the \textit{forced locality} of the exchange potential, which has
the effect that the KS eigenvalues of the occupied states cannot be
associated with ionization energies via Koopmans theorem. In order
to invoke the latter, extra terms similiar in spirit to the derivative
discontinuity of the xc potential have to be considered. Consequently,
even with the exact xc potential the $d$ states cannot be expected
to appear at the experimental positions.

Furthermore, we discussed the EXX(c) electronic structure of the three
cubic perovskites $\mathrm{CaTiO}_{3}$, $\mathrm{SrTiO}_{3}$, and
$\mathrm{BaTiO}_{3}$ in comparison with LDA. The EXX functional opens
the band gap in all materials but leaves the dispersion of the bands
nearly unaffected, except most notably for a slight reduction of the
valence band width. While LDA underestimates the band gaps by about
$50\%$, the EXX(c) potential overcompensates this error and leads
to an overestimation of $25\%$, which we attribute to an incomplete
error cancellation between the neglect of correlation and the neglect
of the xc derivative discontinuity. On the other hand, differences
in the crystal structure between the theoretical setups and experimental
realizations might also play a role.

We note that the IBC is a general approach, which is not restricted
to the FLAPW method. It applies to all electronic structure methods
with a basis set optimized for the effective potential, including
those which are atom-centered and based on pre-calculated and tabulated
basis sets. Typical electronic structure methods of this type include
LMTO,\cite{Local_Orbitals_Andersen,Skriver,Full-potential-LMTO} DMOL,\cite{DMOL}
FHIAIMS,\cite{FHI-AIMS} OPENMX,\cite{OPENMX} and the SIESTA code.\cite{Siesta}

Furthermore, with suitable generalizations the IBC can be applied
to a broad spectrum of response functions in solid-state physics,
e.g., response quantities arising from the displacement of external
potentials (e.g.\ phonons, elastic constants, stress tensor) or from
external fields (e.g.\ g-tensor, chemical shift and nuclear magnetic
resonance, dielectric response, infrared and Raman intensities, magneto-elastic
and magneto-electric tensors), or from more general perturbations
of the Hamiltonian (e.g.\ Born effective charges, polarizability),
to name a few.

Another method that may profit from the IBC is the \textit{GW} approximation
for the electronic self-energy, which involves the calculation of
a dynamical density response function and the single-particle Green
function. It is well known, and Zinc oxide is a prominent example
of it,\cite{ZnO-extreme-I,ZnO-extreme-II} that similarly to the KS
response function in EXX-OEP, \textit{GW} calculations usually converge
badly with respect to the basis set and the number of unoccupied states.
\begin{acknowledgments}
A.G.~gratefully acknowledges funding by the German Research Council
(DFG) through the Cluster of Excellence {}``Engineering of Advanced
Material'' (www.eam.uni-erlangen.de) at the University of Erlangen-Nuremberg.
\end{acknowledgments}
\appendix

\section{Scalar-relativistic equations\label{sec:Scalar-relativistic-equations}}

In the scalar-relativistic approximation\cite{Koelling-Harmon} for
the Dirac equation the large and small radial component, $r^{-1}p(r)$
and $(cr)^{-1}q(r)$, of an electron in a spherical potential $V_{\mathrm{eff},0}(r)$
at energy $\epsilon$ obey the equations of motion\begin{eqnarray}
\frac{dp(r)}{dr} & = & 2m_{\mathrm{rel}}q(r)+\frac{1}{r}p(r)\label{eq: large component p}\\
\frac{dq(r)}{dr} & = & -\frac{1}{r}q(r)+wp(r)\label{eq: small component q}\end{eqnarray}
 with $m_{\mathrm{rel}}=1+[\epsilon-V_{\mathrm{eff,}0}(r)]/(2c^{2})$
($c$ denotes the speed of light) and $w=[l(l+1)]/[2m_{\mathrm{rel}}r^{2}]+V_{\mathrm{eff},0}(r)-\epsilon$.
For a given spherical perturbation that scales with $M(r)$ (index
$I$ omitted to simplify the notation), the linear changes of $p(r)$
and $q(r)$ are given by the solutions of the differential equations\begin{eqnarray}
\frac{dp^{\prime}(r)}{dr} & = & 2m_{\mathrm{rel}}q^{\prime}(r)+\frac{1}{r}p^{\prime}(r)\\
 &  & +\frac{1}{c^{2}}[\epsilon^{\prime}-M(r)]q(r)\nonumber \\
\frac{dq^{\prime}(r)}{dr} & = & -\frac{1}{r}q^{\prime}(r)+wp^{\prime}(r)\\
 &  & -[1+\frac{l(l+1)}{4m_{\mathrm{rel}}^{2}r^{2}c^{2}}][\epsilon^{\prime}-M(r)]p(r)\,.\nonumber \end{eqnarray}
 By direct differentiation we find the differential equations for
the energy derivatives $\dot{p}(r)=dp(r)/d\epsilon$ and $\dot{q}(r)=dq(r)/d\epsilon$
\begin{eqnarray}
\frac{d\dot{p}(r)}{dr} & = & 2m_{\mathrm{rel}}\dot{q}(r)+\frac{1}{r}\dot{p}(r)+\frac{1}{c^{2}}q(r)\\
\frac{d\dot{q}(r)}{dr} & = & -\frac{1}{r}\dot{q}(r)+w\dot{p}(r)\\
 &  & -[1+\frac{l(l+1)}{4m_{\mathrm{rel}}^{2}r^{2}c^{2}}]p(r)\nonumber \end{eqnarray}
 as well as for the linear changes of $\dot{p}(r)$ and $\dot{q}(r)$\begin{eqnarray}
\frac{d\dot{p}^{\prime}(r)}{dr} & = & 2m_{\mathrm{rel}}\dot{q}^{\prime}(r)+\frac{1}{r}\dot{p}^{\prime}(r)\\
 &  & +\frac{1}{c^{2}}\{[\epsilon^{\prime}-M(r)]\dot{q}(r)+q^{\prime}(r)\}\nonumber \end{eqnarray}

\begin{eqnarray}
\frac{d\dot{q}^{\prime}(r)}{dr} & = & -\frac{1}{r}\dot{q}^{\prime}(r)+w\dot{p}^{\prime}(r)\\
 &  & -[1+\frac{l(l+1)}{4m_{\mathrm{rel}}^{2}r^{2}c^{2}}]\nonumber \\
 &  & \times\{[\epsilon^{\prime}-M(r)]\dot{p}(r)+p^{\prime}(r)\}\nonumber \\
 &  & +\frac{2l(l+1)}{8m_{\mathrm{rel}}^{3}r^{2}c^{4}}[\epsilon^{\prime}-M(r)]p(r)\,.\nonumber \end{eqnarray}
 In the nonrelativistic limit $(c\rightarrow\infty)$, these formulas
reduce to Eqs.~\eqref{eq: radial Schroedinger Eq. for u}, \eqref{eq: radial Schroedinger Eq. for u_dot},
\eqref{eq: Sternheimer u}, and \eqref{eq: Sternheimer u dot}.

\bibliographystyle{apsrev4-1} \bibliographystyle{apsrev4-1} \bibliographystyle{apsrev}
\bibliographystyle{apsrev}
\bibliography{biblio}

\begin{thebibliography}{74}
\bibitem[{foo({\natexlab{a}})}]{footnote-I}
\bibinfo{note}{The KS single-particle response function is spin-diagonal for
  the systems considered here. In metallic systems and in the case of
  non-collinear magnetism additional terms can arise that are off-diagonal in
  spin space.}
\bibitem[{foo({\natexlab{b}})}]{footnote-II}
\bibinfo{note}{We note that there is also a corresponding Pulay term for the
  occupied states which is negligible, however.}
\end{thebibliography}

\end{document}